\magnification=\magstep1	
\hsize=14.9truecm	\vsize=8.5truein 
{\dimen0=3truecm \advance \dimen0 by -1truein \global\hoffset=\dimen0}


\font\titlefont=cmb10 at 12pt		\font\authorfont=cmb10
\font\chapterfont=cmb10		        \font\sectionfont=cmti10
\font\symbold=cmbsy10	

\font\vtxt=cmmib10	\font\vscr=cmmib7	\font\vssc=cmmib5
\newfam\vfam		\textfont\vfam=\vtxt 
\scriptfont\vfam=\vscr 	\scriptscriptfont\vfam=\vssc
						

\def\mojdef#1#2#3#4{{\count0=#1 \multiply\count0 by 16
	\advance\count0 by #2	\multiply\count0 by 256
	\advance\count0 by `#4	\global\mathchardef#3=\count0}}
\def\vdef{\mojdef0\vfam}      
\def\bdef{\mojdef0\bffam}


\def\label#1{\global\expandafter\nwcount\csname #1\endcsname}
\catcode`@=11                               	
\def\nwcount{\alloc@@0\count\countdef\insc@unt}
\def\alloc@@#1#2#3#4#5#6#7{\global\advance#7 by1
	\global\advance\count1#1 by 1
	\ch@ck#1#4#2 \allocationnumber=\count1#1
	\global#3#5=\allocationnumber
	\global#5=#7 
	}
\catcode`@=12					


\newcount\enano \enano=0				
\def\eqnm#1 {\label{#1}e\enano \eqno(\equationno{#1})}
\def\eqnmm(#1){\label{#1}e\enano \enspace(\equationno{#1})}
\def\ref(#1){(\equationno{#1})}
\def\refn(#1)#2{(\equationno{#1}#2)}
\def\equationno#1{{\count0=\csname #1\endcsname
	\count2=\count0 \divide\count2 by 100
	\count4=\count2 \multiply\count2 by 100
	\advance\count0 by -\count2 
	\ifnum\count4 < "40 \the\count4.\else 
		{\rm\char\count4}.\fi\the\count0}}

\def\abstract{\vfill\noindent\bf Abstract\endgraf\rm}

\newcount\chapno \chapno=0
\def\chapterno{\chapterfont\the\chapno}
\def\chapter #1\par{\advance\chapno by1 \secno=0 \enano=0 
	\advance\enano by \chapno \multiply\enano by 100 
	\message{[Ch.\the\chapno]} 	\ifvmode\bigbreak\fi 
	\leftline{\chapterno. \chapterfont #1}\nobreak\medskip\rm}
    
\newcount\secno \secno="40
\def\sectionno{\sectionfont\the\chapno.\the\secno.}
\def\section #1\par{\advance\secno by1 \ifvmode\medbreak
	\noindent\fi \leftline{\sectionno\ #1}\nobreak
	\smallskip\rm}

\def\subsection#1 \par{\it #1 \rm}

\newcount\theono \theono=0
\def\theoremno{\chapterno.\the\theono. }
\def\Theorem{\advance\theono by1 \ifvmode\smallbreak
	\noindent\fi \bf Theorem~\theoremno \it }
\def\Proposition{\advance\theono by1 \ifvmode\smallbreak
	\noindent\fi \bf Proposition~\theoremno \it}

\def\title #1\par{\parindent=0pt \null\kern .1in \titlefont #1 \bigskip 
   \authorfont Marijan Ribari\v{c} and Luka \v{S}u\v ster\v si\v c
	\smallskip \rm Jo\v zef Stefan Institute, Jamova 39, p.p.\ 3000, 
		1001 Ljubljana, Slovenia
	\bigskip Corresponding author: Luka \v{S}u\v ster\v si\v c,
		Jo\v zef Stefan Institute,\hfill\break
		Jamova 39, p.p.\ 3000, 1001 Ljubljana, Slovenia\hfill\break
		e-mail: luka.sustersic@ijs.si\hfill\break
		telephone: +386 61 177 3258
		fax: +386 61 123 1569
	\vskip .5in \parindent=20pt \rm}

\def\appendix{\font\sectionfont=cmti10 
   \secno=0 \enano="41 \multiply\enano by 100
   \def\section ##1\par{\advance\secno by1 \ifvmode\medbreak
	\noindent\fi \leftline{\sectionfont A.\the\secno\  ##1}\nobreak
	\smallskip\rm}
    \vfill\eject \leftline{\chapterfont Appendix A} }

\def\heading{\startheading\catcode`\^^M=\active\iftrue\readline}
{\catcode`\^^M=\active							
\gdef\readline#1^^M#2{\fi\headingline{#1 }\ifx#2^^M\endheading	%
	\else \readline#2}}
\def\endheading#1#2#3{\fi\catcode`\^^M=5 \endskip}

\newcount\citno	\citno=0	\def\gobble#1#2{\fi}
\def\decide{\ifx\next\cite ,\else \egroup$\fi}
\def\cite[#1]{\ifmmode \else	$\bgroup \fi
	 \expandafter\the\csname #1\endcsname\futurelet\next\decide}

\def\bibliography{\iftrue\citation}
\def\bswitch{\ifx\next\relax \gobble \else\citation}
\def\citation#1 {\fi \bgroup
	\def\startheading{\label{#1}r\citno}
	\def\headingline##1{} 
	\def\endskip{\egroup\futurelet\next\bswitch} \heading}
\bibliography
Wein4 See, e.g., S.~Weinberg, \it The Quantum Theory of Fields \rm(Cambridge
University Press, Cambridge 1995), Vol.~I, Secs.~12, 1.3, 9.1--9.6, 7.2, 7.3,
and 9; and the references therein. 

Salam A.~Salam, in {\it The Physicist's Conception of Nature}, edited by
J.~Mehra (D.~Reidel, Dordrecht, 1973), p.~430;\\ C.~J.~Isham, A.~Salam and
J.~Strathdee, Phys.\ Rev.\ D \bf 3\rm (1971) 1805; \bf 5\rm (1972) 2548. 

Ross See, e.g., D.~Bailin, Contemp.\ Phys.\ {\bf 30} (1989) 237;\\ D.~J.~Gross,
Nucl.\ Phys.\ B (Proc.\ Suppl.) {\bf15} (1990) 43;\\ G.~G.~Ross, Contemp.\
Phys.\ {\bf 34} (1993) 79;\\ S.~Weinberg, in \it Twentieth Century Physics, \rm
ed. L.~M.~Brown, A.~Pais and B.~Pipard (IOP Publishing and AIP
Publishing, New York, 1995), Vol.~III, p.~2033.

Bjork  J.~D.~Bjorken and S.~D.~Drell, \it Relativistic Quantum 
Fields \rm(McGraw-Hill, New York, 1965), Secs.~11.2 and 11.3;\\ J.~Schwinger,
in \it The Physicist's Conception of Nature, \rm ed. J.~Mehra (D.~Reidel,
Dordrecht, 1973), p.~413.

Heise W.~Heisenberg, Ann.~Phys.~(Leipzig) \bf 32\rm (1938) 20.

Cheng See, e.g., T.~P.~Cheng and L.~F.~Li, \it Gauge Theory of Elementary
Particle Physics, \rm(Claredon Press, Oxford, 1992), Secs.~1.2, 1.1, 5.3, 13.1,
5.1, and~8.1. 

Feynm R.~P.~Feynman, R.~B.~Leighton, and M.~Sands, \it The Feynman Lectures on
Physics \rm(Addison-Wesley, Reading, Mass.,\ 1965), Vol.~II, Sec.~12.7.

Libof See, e.g., R.~L.~Liboff, \it Kinetic Theory \rm(Prentice-Hall, Englewood
Cliffs, N.~J., 1990), Ch.~3, and Sec.~2.4.

Willi See, e.g., M.\ M.\ R.\ Williams, \it Mathematical Methods in Particle 
Transport Theory, \rm(Butterworths, London, 1971), Sec.~2.7 and Ch.~11.

Groot See, e.g., S.~R.~de Groot, W.~A.~van Leeuwen and Ch.~G.~van Weert, 
\it Relativistic Kinetic Theory \rm(North-Holland, Amsterdam, 1980),
Secs.~I.2, VI.1 and VII.

Grad See, e.g., H.~Grad, in \it Application of Nonlinear Partial
Differential Equations in Mathematical Physics, \rm Proc.\ Symp.\ Appl.\ Math.\
{\bf XVII} (Am.\ Math.\ Soc., Providence, R.~I., 1965), p.~154; and
references therein.

Wein1 See, e.g., S.~Weinberg, \it Dreams of a Final Theory \rm(Pantheon Books,
New York 1992).

mi004 M.~Ribari\v c and L.~\v Su\v ster\v si\v c, Transp.\ Theory Stat.\ Phys.\
\bf 24\rm (1995) 1. 

mi002 M.~Ribari\v c and L.~\v Su\v ster\v si\v c, Int.\ J.\ Theor.\ Phys.\ 
\bf 34\rm (1995) 571.

Weidn R.~T.~Weidner, in \it The New Encyclopaedia Britannica \rm (Encyclopaedia
Britannica, Chicago, 1986), 15th edition, Vol.~25, p.845.

Wein2 See, e.g., S.~Weinberg, Phys.\ Rev.\ D\bf7\rm (1973) 1068, Sec.~II;
Rev.\ Mod.\ Phys.\ {\bf 46} (1974) 255, Secs.~I and~III.

Schw1 J.~Schwinger, Phys.\ Rev.\ \bf 91\rm (1953) 713; R.~L.~Arnowitt and
S.~I.~Fickler, Phys.\ Rev.\ \bf 127\rm (1962) 1821.

Fadee See, e.g., L.~D.~Faddeev and A.~A.~Slavnov, \it Gauge Fields:
Introduction to Quantum Theory, \rm(Benjamin Cummings, Reading, Mass., 1980),
2nd edition, Sec.~3.2. 

mi006 For an example see M.~Ribari\v c and L.~\v Su\v ster\v si\v c, Transp.\
Theory Stat.\ Phys.\ \bf 16\rm, 1041 (1987), Secs.~4.5 and 5.1.

Wein3 S.~Weinberg, Ann.\ Phys.\ (N.Y.) {\bf 194} (1989) 336. 

Sudbe For an introduction see, e.g., A.~Sudbery, \it Quantum Mechanics and
Particles of Nature \rm(Cambridge University Press, Cambridge 1988), Chs.~7, 5,
and 2.

Frogg See, e.g., C.~D.~Froggatt and H.~B.~Nielsen, \it Origin of Symmetries,
\rm(World Scientific, Singapore, 1991), and references therein.

McKeo See, e.g., J.~Bernstein, Rev.\ Mod. Phys.\ \bf 46\rm (1974) 7, footnote
32;\\ D.~G.~C.\ McKeon, Can.\ J.\ Phys.\ {\bf 72} (1994) 601.

Grein W.~Greiner, \it Quantum Mechanics, an Introduction, \rm(Springer Verlag,
Berlin, 1989), Sec.~13.2.

Feyng For some related comments see K.~Gottfried and V.~F.~Weisskopf, \it
Concepts of Particle Physics \rm(Claredon Press, Oxford 1984), Vol.~I,
Sec.~13c;\\ R.~P.~Feynman, F.~B.~Morinigo, and W.~G.~Wagner, \it Feynman
Lectures on Gravitation, \rm edited by B.~Hatfield (Addison-Wesley, Reading,
Mass., 1995), Secs~1.4 and 1.5;\\ I.~Percival, Phys.~World \bf 10 \rm(3) 
(1997) 43, and references therein.

mi001 See, e.g., M.~Ribari\v c and L.~\v Su\v ster\v si\v c, Fizika B \bf 3,
\rm (1994) 93; Found.\ Phys.\ Lett.\ {\bf 7} (1994) 531; and references
therein.

Chiao See, e.g., R.~Y.~Chiao, P.~G.~Kwiat, and A.~E.~Steinberg, Sci.\ Am.\ \bf
269 (1993) \rm(2), 38;\\ R.~Y.~Chiao, J.~Boyce, and J.~C.~Garrison, in \it
Fundamental Problems in Quantum Theory, \rm ed. D.~G.~Greenberger and
A.~Zeilinger (The New York Academy of Sciences, New York, 1995), p.~400; and
references therein.

Polya A.~M.~Polyakov, \it Gauge Fields and Strings, \rm(Harwood Academic
Publishers, Chur, 1987), Sec.~1.2.

Kinos T.~Kinoshita and W.~J.~Marciano, in \it Quantum Electrodynamics, \rm
ed. T.~Kinoshita (World Scientific, Singapore, 1990), p.~419.

Brods S.~J.~Brodsky, Preprint SLAC-PUB-95-6781, HEP-PH-9503391 (1995).

\relax

\def\bibliography{\bigbreak
	\leftline{\chapterfont References}\nobreak\medskip 
	\parskip=\smallskipamount \interlinepenalty=10000 
	\parindent=0pt \hangindent=20pt \hangafter=1
	\let\bf=\rm
	\let\it=\rm
	\def\\{\hfill\break}
	\iftrue\citation}
\def\citation#1 {\fi\bgroup  
	\def\startheading{}
	\def\headingline##1{##1}
	\def\endskip{\endgraf\egroup\futurelet\next\bswitch} 
	[\expandafter\the\csname #1\endcsname] \heading}

\def\(#1){\left( #1 \right)}		\def\[#1]{\left[ #1 \right]}

\def\grad{\hbox{\symbold\char'162}}

\def\scpr{\msp{\cdot}\msp} 	 
\def\oot{\msp{\otimes}\msp} 	
\def\msp{\mkern1mu}

\def\OVER#1#2{{\textstyle{#1\over#2}}}   \def\half{\OVER 12}


\def\skupaj #1 {\vskip -#1 \noindent}

\bdef\va{a}   \bdef\gg{g}   \bdef\vp{p}   \bdef\r{r}	\bdef\vx{x}

\vdef\oo\^^a   \vdef\pauv\^^[   \vdef\bb\^^L    \vdef\I{I}   

\def\C{{\cal C}}   \def\P{{\cal P}}
   \def\T{{\cal T}}

\def\L{{\cal L}}   \def\Lfree{\L_{free}}   \def\Lint{\L_{int}}
\def\LNG{\L_I}     \def\LC{\L_{II}} 
\def\LCC{\L_{C}}         
\def\Lq{\L} 	   \def\Lext{\L_{tt}}	   \def\Linf{\L_\infty}

    \def\dol{{\rm L}}     
\def\sA{\fA\kern-.75em{/}\kern.2em}   
\def\sfd{\partial\kern-.75em{/}\kern.2em}	

\def\sta{\Psi}        \def\steq{\sta_{eq}(p;\sta)}
\def\sts{\sta_s}            
\def\stgc{\phi_{gi}}     \def\stg{\sta_g}

\def\stl{\sta_{\as}}   \def\stlz{\sta_{0\as}}   \def\stlh{\sta_{\indfer \as}}  
\def\stlo{\sta_{1\as}}  \def\stlj{\sta_{j\as}}  \def\as{{\rm as}}
\def\Pas{\P_{\as}}

\def\indfer{{{\scriptscriptstyle 1\!{/}2}}}
   
\def\psq{p\scpr p}

\def\argx{(x)}   \def\argp{(p)}   \def\argxp{(x,p)}  \def\argpsq{(\psq)}

\def\brck#1#2#3{\bigl[#1 \,\vert\, #2 \bigr]}
\def\brckz#1#2#3{\bigl[#1 \,\vert\, #2 \bigr](#3)}
\def\intp{\int\!d^4p\,}   

\def\fg{f_g}

\def\Rupshift{\kern-.06em\raise.3ex\null}
\def\RR{{\rm I\kern-.08em R}}	
\def\RRRR{\RR\Rupshift^{1,3}}

\def\ptds{{\textstyle{\partial\over \partial t}}}
\def\ldp{\vbox{\ialign{\hfil$##$\hfil\crcr
   \scriptscriptstyle\leftrightarrow\crcr\noalign{\kern.5pt\nointerlineskip}
   \partial\crcr}}}

\def\scs#1{\phi_{#1}}      \def\scv#1{\phi_{#1}}
\def\ves#1{\varphi_{#1}}   \def\vev#1{\varphi_{#1}}
\def\tes#1{M_{#1}}         \def\tev#1{M_{#1}}

	   \def\fsp{\psi}   
\def\afsp{\overline{\psi}}

\def\sd{p\scpr \partial\kern1.5pt}

\def\Cc#1 #2 #3 { C^{#1}_{#2\,#3}}
\def\CC#1 #2 { C^{#1}_{#2}}
\def\Tr{{\rm Tr}}

\def\lF{L}              \def\lf{\ell}
\def\LF#1{\lF_{#1}}     \def\Lf#1{\lf_{#1}}      \def\Lb{\ell}

\def\fix{\psi}            \def\fip{F}
\def\pfi#1{\fip_{#1}}     \def\xfi#1{\fix_{#1}}  

\def\F#1{{\cal F}_{#1}}

	 	\def\tfin{t_{f}}
\def\metrictensor{\eta}

\overfullrule=0pt

\title Transport-theoretic extensions of quantum field theories

\abstract We propose a new, transport-theoretic (tt) class of relativistic
extensions of quantum field theories of fundamental interactions. Its concepts
are inspired by Feynman's atomistic idea about the physical world and by the
extension of fluid dynamics to shorter distances through the Boltzmann
transport equation. The extending tt Lagrangians imply the original Lagrangians
as path-integralwise approximations. By constructing a tt Lagrangian that
extends a general gauge-invariant Lagrangian, we show that a tt extension of
the standard model is feasible. We define a tt Lagrangian in terms of tt fields
of the spacetime variable and an additional, four-vector variable. We explain
the fields of quantum field theories as certain covariant, local averages of tt
fields. Only two tt fields may be needed for modeling fundamental interactions:
(i)~a four-vector one unifying all fundamental forces, and (ii)~a
two-component-spinor one unifying all fundamental matter particles. We comment
on the new physics expected within the tt framework put forward, and point out
some open questions. 
\vfill
\leftline{PACS: 11.10.Kk, 12.10.Dm}
\leftline{Keywords: Extension, quantum field theory}
\eject

\chapter Introduction and motivation

One can describe the strong, weak, and electromagnetic interactions
at presently accessible energies, with required precision, by the
standard model. Nowadays, however, this model is generally believed
to be a low-energy approximation to a more fundamental theory
[\cite[Wein4]]. Yet it is not clear to what kind of theory, and what
physics is lost as a result of this approximation; e.g., Salam
[\cite[Salam]] suggested that the physics of quantum gravity is lost.
Some theorists are looking for an improved theory of fundamental
interactions outside the framework of quantum field theories in
four-dimensional spacetime $\RRRR$ (QFTs), formulated in terms of
fields of only four independent, continuous variables. To explain
premises of the standard model, to extend it, and to include gravity,
they study quantum field theories in higher-dimensional spacetimes
$\RR\Rupshift^{1, n}$, $n > 3$ [\cite[Ross]].

The ultraviolet divergences of realistic QFTs may be seen as a sign of
inadequate treatment of physical processes at higher energies---processes
determined by the experimentally unexplored, small-distance physics of
fundamental interactions [\cite[Wein4]\cite[Bjork]]. Already sixty years ago,
Heisenberg [\cite[Heise]] proposed that a QFT can provide only an idealized,
large-scale description of quantum dynamics, valid for distances larger than
some fundamental length. As it is still not clear how to extend present QFTs of
fundamental interactions to shorter distances, we will consider to this end a
new quantum field theory in eight dimensional $\RRRR \times \RRRR$.

It is nowadays customary to specify a QFT by choosing (1)~a {\it finite number}
of $c$-number fields of the spacetime variable $x \in \RRRR$ (fields), $\xfi k
\argx$, and (2)~a classical Lagrangian density (Lagrangian) $\L$ defined in
terms of $\xfi k$ and of their first-order, spacetime derivatives $\partial_\mu
\xfi k$. The QFT probability amplitudes, representing quantum dynamics, can
then be expressed in terms of the Feynman path integral
[\cite[Wein4]\cite[Cheng]]
$$   \int [ d\xfi k] \exp \bigl( (i/\hbar) I \bigr) \,,    \eqnm Fptin $$
where the action
$$   I[\dots,\xfi k,\dots] \equiv c^{-1} \int \L \, d^4 x \,.
							   \eqnm actio $$
To extend a QFT, we are going to look for such an extending Lagrangian that
the path integrals specified by it may be approximated by the corresponding QFT
path integrals, providing some characteristic length of the extending
Lagrangian tends to zero.

The Euler-Lagrange equations
$$   {\partial\L\over \partial\xfi k} - \partial_\mu\Bigl( {\partial\L
        \over \partial (\partial_\mu\xfi k) } \Bigr) = 0 
	   						   \eqnm EuLeq $$
of present QFTs are nonlinear, coupled, covariant, partial-differential
equations [\cite[Wein4]\cite[Bjork]\cite[Cheng]]. Following Bjorken and Drell
[\cite[Bjork]] and Feynman [\cite[Feynm]], we may wonder once again, some
thirty years later, what kind of equations could be the Euler-Lagrange
equations in theories that will extend present QFTs to shorter distances?
Commenting on the ``underlying unity'' of nature, Feynman [\cite[Feynm]] noted
that the partial-differential equation of motion ``\dots we found for neutron
diffusion is only an approximation that is good when the distance over which we
are looking is large compared with the mean free path. If we looked more
closely, we would see individual neutrons running around.'' And then he
wondered, ``Could it be that the real world consists of little X-ons which can
be seen only at very tiny distances? And that in our measurements we are always
observing on such a large scale that we can't see these little X-ons, and that
is why we get the differential equations? \dots Are they [therefore] also
correct only as a smoothed-out imitation of a really much more complicated
microscopic world?'' Presuming that Feynman was on the right track with this
atomistic idea, we believe that the kinetic theory of gases suggests a way to
obtain such Lagrangians for modeling fundamental interactions that extend the
Lagrangians of QFTs to shorter distances.

In the kinetic theory of gases, one describes large-scale phenomena by a {\it
finite number} of fields, so-called macroscopic variables such as number
density, macroscopic mean velocity, pressure tensor, heat-flow vector, and
kinetic-energy density [\cite[Libof]\cite[Willi]]. For such large-scale
phenomena where changes in macroscopic variables over a mean collision time and
over a mean free path are sufficiently small, one can predict the evolution of
these fields accurately enough by the partial-differential equations {\it of
fluid dynamics}. Otherwise, one must resort to a more detailed description of
physical processes. To better approximate the evolution of macroscopic
variables, one may use the one-particle distribution: a function of spacetime
positions $x$ and of four-momenta $p$ of constituent particles, i.e., of eight
independent, continuous variables. This classical function of $x, p \in \RRRR$
(tt field) determines the values of macroscopic variables through certain local
averages over four-momenta $p$. For a sufficiently rarefied gas, it is accepted
that the equation of motion for such tt field is an integro-differential,
Boltzmann transport equation, which is local in the spacetime variable $x$, but
not in the four-momentum variable $p$. As it takes some account of gas
``granularity'', the Boltzmann equation describes dynamics of physical
processes also at vastly smaller distances than the partial-differential
equations of fluid dynamics. These small-scale physical processes determine the
large-scale, fluid phenomena described by macroscopic variables. The natural
time of evolution of macroscopic variables, which can be adequately determined
by the equations of motion of fluid dynamics, apparently bears no relation to
the mean collision time, a vastly smaller, basic time-scale for the
corresponding solution to the Boltzmann equation [\cite[Grad]]. Within the
transport-theoretic (tt) framework, this disparity causes no conceptual
problems such as the hierarchy problem in elementary particle physics
[\cite[Wein1]].

One can formally derive the equations of motion of fluid dynamics as
such an asymptotic approximation to the Boltzmann equation that
depends on its initial and boundary conditions [\cite[Grad]]. Which
explains fluid dynamics as an asymptotic theory in the kinetic theory
of gases. Inspired by this explanation, and by Feynman's conjecture
that the partial-differential equations of QFTs are actually
describing large-scale phenomena due to microscopic motion
of some hypothetical X-ons, we formulated [\cite[mi004]] certain new,
relativistic, integro-differential, transport equations:
equations which offer such physically motivated, tt extensions of
partial-differential equations of a QFT to shorter distances that
contain them as large-distance approximations. When thinking about
alternatives to the conventional QFTs, such extensions were found
missing by Bjorken and Drell [\cite[Bjork]] on the analogy with the
classical, partial-differential equations that are an idealization
valid for distances larger than the characteristic length measuring
the granularity of the medium. So we proposed [\cite[mi002]] that for
modeling quantum dynamics of fundamental interactions by the
path-integral method we use, instead of QFT Lagrangians, such tt
Lagrangians that imply them as asymptotic approximations, and whose
Euler-Lagrange equations are integro-differential, transport
equations. Here, our aims are: (1)~to present the basic concepts of
relativistic, tt quantum field theories (ttQFTs) specified by such tt
Lagrangians, (2)~to formally show that ttQFTs extending QFTs of
fundamental interactions are feasible, (3)~to elaborate on physical
implications of ttQFTs, and (4)~to point out relevant physical
questions.

The paper is organized as follows: In Section~2, we specify basic
properties of tt fields, and of tt Lagrangians extending QFT
Lagrangians. We introduce a length $\lambda$ to classify present QFTs
as asymptotic approximations to certain ttQFTs as $\lambda \to 0$. In
Section~3, we give a covariant, tt Lagrangian that extends a
general gauge-invariant Lagrangian. In Section~4, we give tt
symmetry transformations generalizing those of QFTs. In Section~5, we
point out what kind of physics might distinguish an extending ttQFT
from the original QFT. We also consider the transition from a ttQFT
to the corresponding QFT. Concluding remarks and summary are given in
Sec.~6.  Mathematical formalism of ttQFTs is considered in the
Appendix, where we give simple tt Lagrangians extending the
Klein-Gordon, Dirac, and QED Lagrangians.  For convenience, we use
the metric tensor $\metrictensor^{\mu \nu} \equiv {\rm diag}( -1, 1,
1, 1)$, bispinors and Dirac matrices in the chiral representation,
and the natural units in which $\hbar = c = 1$; $[\dol]$ denotes
dimensions of length and time, and $[\dol]^{-1}$ those of energy,
momentum, and mass.

\chapter Basic assumptions of tt extensions

In a ttQFT that extends a QFT, we will assume that one can study quantum
processes modeled by this QFT using (1)~tt fields of two independent,
four-vector variables; (2)~a tt Lagrangian defined in terms of tt fields and 
their spacetime derivatives; and (3)~appropriate path integrals, specified by
this tt Lagrangian.

\section Transport-theoretic fields

We will specify a tt Lagrangian in terms of real and complex, relativistic tt
fields, say, $\sta_j\argxp$ of the spacetime variable $x = (t,\r) \in \RRRR$
and of an additional, four-vector variable $p = (p^0,\vp) \in \RRRR$. The
number and type of tt fields $\sta_j\argxp$ we will use depend on the physical
system under consideration, the interactions we take account of, and the way we
model them. Dimensions of the independent variable $x$ are $[\dol]$; for 
convenience, the dimensions of the independent variable $p$ and of tt fields
$\sta_j\argxp$ are chosen to be $[\dol]^{-1}$ and $[\dol]$, respectively. In 
contrast to field theories in higher-dimensional spacetimes, in ttQFTs the
additional independent variables $p^0$, $p^1$, $p^2$ and $p^3$ (which we may
regard as the components of four-momenta of Feynman's X-ons) do not make us
modify our ideas about time and space. In ttQFTs, the spacetime is still the
same as in QFTs, the four-dimensional space $\RRRR$, and no compactification is
necessary.

We will need Lorentz-scalar (scalar) tt fields $\sta_0\argxp$,
left-handed--two-com\-po\-nent-spinor (left-spinor) tt fields $\sta_{\indfer}
\argxp$, right-handed--two-component-spinor (right-spinor) tt fields
$\sta_{-\indfer}\argxp$, and four-vector tt fields $\sta_1\argxp$.
On the analogy of the relativistic kinetic theory, we assume that under an
inhomogeneous, proper, ortochronous, Lorentz transformation $x \to \Lambda x +
a$, $a \in \RRRR$, a tt field
$$   \sta_j\argxp \to U(\Lambda, a) \sta_j\argxp \equiv 
        D_j(\Lambda^{-1}) \sta_j \bigl( \Lambda x + a, 
	\Lambda p \bigr) \,,                               \eqnm Lortr $$ 
where $D_j(\Lambda)$, $j = 0$, $1/2$, $-1/2$, $1$, are the conventional
Lorentz-transformation matrices for scalars, left-spinors, right-spinors, and 
four-vectors, respectively. [$D_0 \equiv 1$; $D_\indfer^\dagger = D_{-\indfer
}^{-1}$; $D_1 \equiv \Lambda$; for the four-vector tt field $\sta_1 \argxp
\equiv p$, $U(\Lambda, a)\sta_1 = p$.] Under Lorentz transformations
\ref(Lortr) the independent variables $x$ and $p$ do not mix; $x$ transforms as
the spacetime four-vector, and $p$ transforms as a momentum four-vector. So no
extension of special relativity is needed in ttQFTs.

To define free tt Lagrangians we will use the following local, bilinear
mappings $\brck{\sta_j}{\sta_j'}{x}$ of tt fields $\sta_j\argxp$ and
$\sta'_j\argxp$ into scalar fields:
$$   \eqalignno{
     \brckz{\sta_0}{\sta_0'}{x} &\equiv \intp \sta_0^*(x,-p) 
        \sta_0'\argxp \,,                          \cr
     \brckz{\sta_{\pm\indfer}}{\sta_{\pm\indfer}'}{x} &\equiv 
        \intp \sta_{\pm\indfer}^\dagger (x,-p) \sigma_\pm\argp 
        \sta_{\pm\indfer}' \argxp \,,       &\eqnmm(scprd)\cr
     \brckz{\sta_1}{\sta_1'}{x} &\equiv \intp \sta_1^*(x,-p) 
        \scpr \sta_1'\argxp \,.                    \cr}      $$
In the above: ${}^*$ denotes complex conjugation; ${}^\dagger$ denotes the
adjoint (i.e., transposed and complex-conjugate) spinor; the $2 \times 2$
matrix functions 
$$   \sigma_\pm\argp \equiv i(\psq)^{-1/2} \bigl( \vp\scpr \pauv \pm
        p^0 I \bigr)                                       \eqnm paufv $$
for $\psq > 0$, where $I$ is the $2 \times 2$ identity matrix, and $\pauv =
(\sigma^1, \sigma^2, \sigma^3)$ is the vector of Pauli matrices;  $a\scpr b
\equiv a^\mu b_\mu$, $a$, $b \in \RRRR$; for a function $F(p)$ of $p \in
\RRRR$, the functional
$$   \intp F(p) \equiv -i \lim_{R \to \infty} \int _{-iR}^{iR} dp^0 
        \int_{\psq < R^2} F(p) \,d^3\vp \,,                \eqnm intpd $$
where $F\argp$ is defined for imaginary values of $p^0$ through analytic
continuation. Henceforth, we assume that tt fields $\sta_j\argxp$ are such
that all functionals \ref(intpd) we will encounter exist and are invariant in
the sense that $\intp F(\Lambda p) = \intp F\argp$, see [\cite[mi004]]. $\intp$
is an analog of the standard Euclidean definition of a fourfold integral
through Wick's rotation and symmetric integration. 

Mappings $\brck{\sta_j}{ \sta_j'}{x}$ are scalar fields in the sense that
[\cite[mi004]] 
$$   \brckz{U(\Lambda,a)\sta_j}{ U(\Lambda,a)\sta'_j}{x} = \brckz{\sta_j}{
        \sta_j'}{\Lambda x + a} \,.           \eqnm fLLin $$
They do not explicitly depend on $x$, and are strictly local in the sense that
they ignore the $x$-dependence of tt fields: 
$$   \brckz{\sta_j}{ \sta'_j}{a} = \brckz{\sta_j(a,p)}{
        \sta_j'(a,p)}{a} \,.			           \eqnm fLLlo $$

\section Lagrangian of a ttQFT

We denote by $\sd$ the covariant, substantial time derivative $p^0\ptds +
\vp\scpr\grad$, by $\sta\argxp$ the tt field $(\ldots, \sta_j\argxp, \ldots)$
consisting of all tt fields $\sta_j\argxp$ of the considered ttQFT, and by
$\Lext(\sta, \sd\sta)$ the tt Lagrangian defining this ttQFT. We write a tt
Lagrangian as a sum of free and interaction parts,
$$  \Lext = \Lfree + \Lint \,,                             \eqnm Lgsum $$
which are real, scalar fields in the sense of \ref(fLLin); do not explicitly
depend on the spacetime variable $x$; and have dimensions of energy density,
$[\dol]^{-4}$. Thus the tt action $I[\sta] \equiv \int \Lext \,d^4 x$ is a
real, dimensionless, relativistic invariant.

The {\it free part} of a tt Lagrangian
$$   \Lfree \equiv \half \sum_j \bigl\{ \brck{\sta_j}{\sd\sta_j}{x} +
         \brck{\sd\sta_j}{\sta_j}{x} \bigr\}\,, 	   \eqnm Lfdef $$
where $\brck{\sta_j}{\sd\sta_j}{x}$ and $\brck{\sd\sta_j}{\sta_j}{x}$, $j=0,
1$, are equal and real if $\sta_j$ is real. The Euler-Lagrange equations of
$\Lfree$ are the covariant, partial-differential equations 
$$   \sd \sta_j\argxp = 0 \,,                              \eqnm freqm $$
by (A.7). [For solutions to \ref(freqm) see [\cite[mi002]]. For an alternative
to definition \ref(Lfdef) see (A.27).] According to \ref(freqm), all
Euler-Lagrange equations of free tt Lagrangians are of the same simple form:
the tt approach entails certain unification and simplification of free
Lagrangians. On the analogy of the kinetic theory of gases, we may consider
equations of motion \ref(freqm) for free tt fields as a classical description
of free streaming of Feynman's X-ons: an analog of Newton's first law.

We assume that the {\it interaction part} $\Lint$ of a tt Lagrangian acts
strictly locally on tt fields $\sta\argxp$, i.e., satisfies a relation such as
\ref(fLLlo), and so does not depend on the spacetime derivatives $\partial\sta$
of $\sta$. If regarded as a part of a classical tt description of Feynman's 
X-ons, the interaction part $\Lint(\sta)$ of a tt Lagrangian $\Lext$ describes
strictly local interactions of infinitesimal entities: an example of the
Aristotle concept of contact forces.

\section Extension of a QFT

To be an extension of certain QFT, a ttQFT has to respect the
correspondence principle; i.e., as pointed out by Weidner
[\cite[Weidn]]: ``\dots when a more general theory is applied to a
situation in which a less general theory is known to apply, the more
general theory must yield the same predictions as the less general
theory.'' So, when calculating the ttQFT transition amplitude for a
quantum process that can be adequately described already by the QFT,
we should be able to replace the ttQFT path integral, specified by
the tt Lagrangian $\Lext(\sta, \sd\sta)$ in terms of functions
$\sta\argxp$ of {\it eight} independent variables, with the
corresponding QFT path integral, specified by the QFT Lagrangian
$\Lq(\ldots, \xfi k, \partial \xfi k, \ldots)$ in terms of functions
$\xfi{k} \argx$ of only {\it four} independent variables.  Let us
show how such a replacement of ttQFT with the QFT can be explained.

Inspired by the explanation of fluid dynamics as an approximate theory in the
kinetic theory of gases [\cite[Libof]--\cite[Grad]], we introduce the set of
asymptotic tt fields
$$   \stl\argxp \equiv \sum_k \pfi{k}\argp \xfi{k}\argx \,.
							   \eqnm stjas $$
They are sums of appropriate products of the QFT fields $\xfi{k}\argx$ and of
functions $\pfi{k}\argp$ of $p \in \RRRR$ that are the same for all
$\stl\argxp$, though they may depend on $\Lext$ but (to simplify) not on the
initial and final states [$\pfi k\argp$ will play here a role analogous to that
of the Maxwellian distribution and corrections to it.]. We then hypothesize
that the quantum process for which a ttQFT transition amplitude is adequately
approximated by the QFT one is in general such that:

(a)~We can specify the initial and final states of this process by the QFT
fields $\xfi k\argx$.

(b)~We can obtain an adequate approximation to the ttQFT path integral
specified by the tt Lagrangian $\Lext$ and the initial and final states by
integrating over all such asymptotic tt fields $\stl\argxp$ whose fields $\xfi
k$ are constrained by the initial and final states {\it in the same way} as in
the corresponding QFT path integral specified by $\L$. Thus the actual
variables of the ttQFT path integral over such asymptotic tt fields are the
same as in the corresponding QFT path integral: the fields $\xfi k\argx$.

(c)~These two path integrals are equivalent.

Whenever these three conditions are met, we will regard (i)~the Lagrangian
$\Lq$ of the original QFT as a path-integralwise approximation to the tt
Lagrangian $\Lext$, and (ii)~the tt Lagrangian $\Lext$ as a tt extension of the
QFT Lagrangian $\Lq$ provided $\Lext$ with $\sta$ restricted to $\stl$ equals
$\Lq$, i.e.,
$$   \Lext(\stl, \sd\stl) = \Lq(\ldots, \xfi{k},
        \partial\xfi{k}, \ldots ) \,.			   \eqnm Lttft $$

In the classical asymptote $\hbar \to 0$, the formal oscillatory behaviour of
the integrand of \ref(Fptin) suggests the main contribution to the Feynman path
integral is due to the fields $\xfi k\argx$ that satisfy the boundary
conditions imposed by the initial and final states and make the action $I$
stationary by satisfying the Euler-Lagrange equations \ref(EuLeq). On the
analogy of this formal asymptote in QFTs, in this paper we will formally
simulate such quantum processes where the QFT is adequate by replacing the
above condition (b) with the following assumptions: (1)~The interaction part
$\Lint$ of the tt Lagrangian $\Lext$ {\it depends on a characteristic length}
$\lambda$, $[\lambda] = [\dol]$, and, in general, tends to infinity if $\lambda
\to 0$. (This length $\lambda$ is an analog of the singular parameter
introduced in 1924 by Hilbert to formally obtain fluid dynamics as an
asymptotic approximation in the kinetic theory of gases [\cite[Grad]].) (2)~The
asymptotic tt fields \ref(stjas) are the most general tt fields that satisfy
the Euler-Lagrange equations of $\Lext$ in the leading order of $\lambda$.
(3)~We are considering such transitions where $\lambda$ is relatively so small
that we can obtain adequate approximations to the ttQFT path integrals by
integrating over the asymptotic tt fields $\stl\argxp$ with fields $\xfi
k\argx$ consistent with the initial and final states. We will refer to such
asymptote $\lambda \to 0$ as the QFT asymptote since we assumed that eventually
as $\lambda \to 0$: (i)~the tt Lagrangian $\Lext$ actually depends only on the
QFT fields $\xfi{k} \argx$, and (ii)~tt transition amplitudes can be expressed
in terms of the path integrals specified by the QFT Lagrangian $\Lq$ (which may
depend on $\lambda$). In Sec.~5.2, we will mention how quantum processes one
models by a QFT may come about.

Henceforth, we assume that functions $\pfi k\argp$ are such that the
incorporation \ref(stjas) of QFT fields $\xfi k\argx$ into the asymptotic tt
fields $\stl\argxp$, the $p$-dependence of asymptotic tt fields $\stl\argxp$,
and the definition \ref(Lttft) of a tt extension of a QFT Lagrangian $\L$ are
relativistically invariant in the sense that
$$    \eqalignno{
	U(\Lambda, a)\stl\argxp &= \sum_k \pfi{k} \argp U(\Lambda, a)
		\xfi{k}\argx \,, 		&\eqnmm(stlco)\cr
	\Lext(U(\Lambda, a)\stl, \sd U(\Lambda,a)\stl) &= 
		\L(\ldots, U(\Lambda,a)\xfi k, \partial
		U(\Lambda,a)\xfi k,\ldots) \,.	&\eqnmm(extco)\cr}	$$
By \ref(stlco), the asymptotic tt fields \ref(stjas) with the
Lorentz-transformed QFT fields equal the Lorentz-transformed asymptotic tt
fields.

\section Macroscopic variables

Take a ttQFT that extends certain QFT. Motivated by the macroscopic
variables in the kinetic theory of gases, we assume that for each kind of QFT
fields $\xfi{k}\argx$ there is a special covariant average of any ttQFT field
$\sta\argxp$ over the independent variable $p$, namely, the tt macroscopic
variable
$$   \xfi{k}[x; \sta] \equiv \intp \F k\argp \sta\argxp    \eqnm macva $$
of $\sta\argxp$, where $\F{k}\argp$ is such a function of $p \in \RRRR$ that
$\xfi{k}[x; \sta]$ and $\xfi{k}\argx$ are the same kind of relativistic fields,
have the same dimensions, satisfy relations 
$$    \xfi{k}[x; \stl] 	= \xfi{k}\argx 			   \eqnm macas $$
and
$$    \xfi{k}[x; U(\Lambda,a) \sta] = U(\Lambda,a) \xfi{k}[x; \sta] \,,
							   \eqnm macLt $$
and {\it have the same physical significance} in the following sense: If a tt
field $\sta\argxp$ is consistent with certain initial and final states, we can
take its macroscopic variables $\xfi{k}[ x;\sta]$ as a QFT path between these
two states. Hence, the initial and final states constrain the macroscopic
variables $\xfi k[x; \sta]$ of ttQFT fields $\sta\argxp$ in the same way as the
QFT fields $\xfi k \argx$. So we may generalize \ref(macas) and regard QFT
fields as the macroscopic variables of the ttQFT extension and vice versa; this
connection between ttQFT and QFT fields is relativistically invariant, by
\ref(macLt). By \ref(macas) and \ref(stjas), an asymptotic tt field is
completely described by its macroscopic variables:
$$   \stl\argxp = \sum_k \pfi{k}\argp \xfi{k}[x;\stl] \,. 	
							   \eqnm Stjas $$

Using macroscopic variables $\xfi k[x;\sta]$ and the characteristic functions
$\pfi k\argp$ appearing in \ref(stjas), we define the macroscopic projection
operator 
$$	\Pas \sta \equiv \sum_k \pfi k\argp \xfi k[x;\sta] \,;
							   \eqnm macpr $$
$\Pas^2 \sta = \Pas \sta$, $\Pas \stl = \stl$, and $\xfi k[x; \Pas\sta] = \xfi
k[x; \sta]$, by \ref(stjas) and \ref(macas). If $\sta$ is a tt path, then
$\Pas\sta$ is an asymptotic tt field \ref(stjas) with fields $\xfi k$
compatible with initial and final states. Thus any ttQFT path $\sta\argxp$
consistent with certain initial and final states can be written as a sum
$$	\sta\argxp = \stl\argxp + (1 - \Pas) \sta\argxp \,,   \eqnm cdsta $$
where $\stl$ is defined by \ref(stjas) with $(\ldots,\xfi k\argx,\ldots)$ being
a path between these two states in the original QFT. Which shows that extending
a QFT by a ttQFT, we actually add tt fields $(1 - \Pas)\sta$ to asymptotic
tt fields, so to speak; so an infinite number of ttQFT paths corresponds to one
QFT path. In the QFT asymptote, the part $(1 - \Pas) \sta$ of any ttQFT path
$\sta$ is ``neglected''. We have yet to find out how the parts $(1 - \Pas)
\sta$ of ttQFT paths $\sta$ are constrained by the initial and final states,
and by the ttQFT Lagrangian $\Lext$. If the initial and final values of $(1 -
\Pas) \sta$ are uniquely determined, the extending ttQFT implies no new
particles in addition to those considered in the original QFT. And if they are
equal to zero, the asymptotic tt fields $\stl$ are ttQFT paths.

In Section~3 and in the Appendix, we are going to consider tt extensions of
QFT Lagrangians. In this paper, we will not consider: (1)~how to define the
macroscopic variables and tt path integral appropriate to a given tt Lagrangian
and certain initial and final states; (2)~the quantization and derivation of
Feynman's rules in ttQFTs; (3)~which specific gauge and ghost terms, and
regularization and renormalization procedures, a ttQFT implies in the QFT it
extends; and (4)~how a tt extension modifies the divergent integrals of a QFT.

\chapter Transport-theoretic extensions of a QFT Lagrangian

In this Section, we are going to look into the physical, conceptual problems in
constructing a tt extension of a QFT Lagrangian. To this end, we will make the
following basic assumptions considered in Sec.~2: (1)~A tt Lagrangian is
defined in terms of tt fields. It is a sum \ref(Lgsum) of free and interaction
parts, which are real, scalar fields and do not depend explicitly on $x$.
(2)~The free part of a tt Lagrangian is given by \ref(Lfdef). (3)~The
interaction part of a tt Lagrangian is local and without derivative
couplings. It depends on a characteristic length $\lambda$ and, in
general, tends to infinity if $\lambda \to 0$. (4)~When calculating
the ttQFT probabilities amplitudes for quantum processes that are
adequately modeled already by the QFT to be extended, we may take as
tt paths the asymptotic tt fields \ref(stjas) that satisfy the tt
Euler-Lagrange equations in the leading order in $\lambda$ and whose
fields $\xfi k\argx$ are compatible with the initial and final
states. (6)~So computed ttQFT probability amplitudes equal the
corresponding QFT ones.

Step by step to ever higher energies, it took some fifty years of intense
theoretical and experimental effort to establish first the QED and then the
standard model. To show that a bottom-up tt approach to fundamental
interactions that takes account of these results is feasible, we are going to
construct a ttQFT formally encompassing the standard model. To this end, we
start with the general gauge-invariant Lagrangian [\cite[Wein2]]
$$   \eqalignno{
     \LC &\equiv - \OVER 14 \bigl( \partial_\mu A_{a\nu} - 
        \partial_\nu A_{a\mu} - C_{abc} A_{b\mu} A_{c\nu} \bigr)
        \bigl( \partial^\mu A^\nu_a - \partial^\nu A^\mu_a 
	- C_{ade} A^\mu_d A^\nu_e \bigr)       \cr
     &\qquad{}- \afsp_m \bigl[ \gamma^\mu \partial_\mu \fsp_m + 
        ( i \gamma^\mu A_{a\mu} T^a_{mn} + m_{mn}
	+ \phi_i \Gamma^i_{mn} ) \fsp_n \bigr]   &\eqnmm(nAcdf)\cr
     &\qquad{}- \half (\partial^\mu \phi_i + i A^\mu_a R^a_{ij} \phi_j )
        ( \partial_\mu \phi_i + i A_{b\mu} R^b_{ik} \phi_k )
        + \half\mu^2 \phi_i \phi_i - \OVER 14 \mu^2 v^{-2}
        (\phi_i \phi_i)^2 \,.                      \cr}                   $$
In the above: (1)~$A_a\argx$ are real, four-vector, gauge fields;
$\fsp_m\argx$ are complex, chiral-bispinor fields; and $\phi_i\argx$
are real, scalar fields.  (2)~$\afsp_m\argx \equiv \fsp^\dagger_m
\argx \beta$, with $\fsp_m^\dagger$ being the bispinor adjoint to
$\fsp_m$ and $\beta = i\gamma^0$; and $\gamma^\mu$ are the four $4
\times 4$ Dirac matrices. (3)~$T^a_{mn}$ and $R^a_{ik}$ are
components of Hermitian matrices representing the Lie algebra of the
gauge group of $\LC$ ($iR^a_{ik}$ are real). $T^a_{mn}$, $R^a_{ik}$,
and the structure constants $C_{abc}$ are proportional to one or more
gauge coupling constants.  (4)~$\Gamma^i_{mn}$ are components of
Yukawa coupling matrices; $m_{mn}$ are components of a bare-mass
matrix; and $\mu^2$ and $v^2 \ne 0$ are real constants. (5)~Here and
in what follows, the repeated, Lorentz or gauge indices are summed
over.

Lagrangian $\LC$ itself does not have a tt extension as it contains the
first-order derivatives also quadratically and in its interaction part,
cf.~\ref(Lfdef). We therefore take an equivalent first-order Lagrangian, the
real scalar field
$$   \eqalignno{
     \LNG &\equiv \LF{ab} F_b^{\mu\nu} \bigl[ \ldp_\nu A_{a\mu} 
        + C_{acd} A_{c\mu} A_{d\nu} 
	+ \LF{ac} F_{c\mu\nu} \bigr]             \cr
     &\qquad{}- \afsp_m \bigl[ \OVER{1}2 \gamma^\mu \ldp_\mu \fsp_m + 
        (i \gamma^\mu A_{a\mu} T^a_{mn} + m_{mn}
	+ \phi_i \Gamma^i_{mn} ) \fsp_n \bigr]   \cr
     &\qquad{}+ \half\Lf{ij} \varphi_j^\mu \bigl[ 
        \ldp_\mu \phi_i + 2i A_{a\mu} R^a_{ik} \phi_k +
	\Lf{ik} \varphi_{k\mu} \bigr]            &\eqnmm(nAlin)\cr
     &\qquad{}+ \half\mu^2 \phi_i \phi_i - \mu^2 v^{-2} \Lb b 
        \phi_i \phi_i + \mu^2 v^{-2} \Lb^2 b^2 \,,      \cr}      $$
where: (1)~$A_a$, $\fsp_n$, and $\phi_n$ are the fields of $\LC$;
$F_a\argx$ are real, antisymmetric-tensor fields; $\varphi_i\argx$
are real, four-vector fields; and $b\argx$ is a real, scalar field.
(2)~Real, dimensionless parameters $\LF{ab}$ and $\Lf{ij}$ are
elements of two invertible matrices, and $\Lb \ne 0$ is a real,
dimensionless parameter. (3)~$a\ldp_\mu b\equiv a (\partial_\mu b) -
(\partial_\mu a) b$. Within the framework of QFTs, Lagrangian $\LNG$
is equivalent to $\LC$ since their Euler-Lagrange equations are
equivalent [\cite[Schw1]]. Physical significance of parameters
$\LF{ab}$, $\Lf{ij}$, and $\Lb$ remains to be found out. If $\LF{ab}
=\delta_{ab}$, $\Lf{ij} = \delta_{ij}$, and $\Lb = 1$, the Lagrangian
$\LNG$ is actually, up to a divergence term, a general
gauge-invariant Lagrangian in the first-order formalism
[\cite[Fadee]]. Hence $\LNG$ is also gauge-invariant (up to a
divergence term) since $\LF{ab}$ and $\Lf{ij}$ are elements of
invertible matrices.

Unlike the conventional Lagrangian $\LC$ of non-Abelian gauge
theories, the equivalent first-order Lagrangian $\LNG$ has already
two basic properties of tt Lagrangians: Owing to the additional
fields $F_a\argx$ and $\varphi_i\argx$, the first-order derivatives
appear only linearly, and there are no derivative couplings, i.e., no
interaction terms involving derivatives. Lagrangian $\LNG$ also has
no quartic interaction terms; to this end we introduced the scalar
field $b\argx$, which is therefore not needed if $v^{-2} = 0$ in
$\LC$.

\section A tt extension of a non-Abelian--gauge theory

To construct a ttQFT Lagrangian that formally extends the QFT Lagrangian
$\LNG$, we take the tt field $\sta = (\sta_0, \sta_\indfer, \sta_1)$ with real
$\sta_0$ and $\sta_1$ but complex $\sta_\indfer$. And we incorporate the fields
$\phi_i \argx$, $\varphi_i\argx$, $b\argx$, $\fsp_m\argx$, $A_a\argx$, and
$F_a\argx$ of $\LNG$ in the asymptotic tt field $\stl = (\stlz, \stlh, \stlo)$
as follows:
$$   \eqalign{
     \stlz\argxp &\equiv f_{0i}\argpsq \phi_i\argx + 2f_{1i}\argpsq 
        \varphi_i \argx \scpr p + f_2\argpsq b\argx \,,    \cr     
     \stlh\argxp &\equiv \bigl( \sigma_-\argp, I \bigr) f_{\indfer m}
        \argpsq \fsp_m\argx \,,                  \cr
     \stlo\argxp &\equiv (1 - \P_l) f_{3a}\argpsq A_a\argx 
	+ 2 f_{4a}\argpsq F_a\argx p \,.         \cr}      \eqnm asnAs $$
Here: $\bigl( \sigma_-\argp, I \bigr)$ is a $4 \times 2$ matrix; $(Ta)^\mu
\equiv T^{\mu\nu} a_\nu$ for a second-rank tensor $T$ and a four-vector $a$;
the longitudinal projection $\P_l a \equiv \argpsq^{-1} (p\scpr a) p$ for a
four-vector $a$; and $f_{0i}\argpsq$, $f_{1i}\argpsq$, $f_2\argpsq$,
$f_{\indfer m} \argpsq$, $f_{3a}\argpsq$, and $f_{4a}\argpsq$ are such real
functions of $\psq > 0$ that the real matrices $\CC 1 00 $, $\CC 2 11 $, $\CC 1
{\indfer\,\indfer} $, $\CC 1 33 $, $\CC 2 44 $, $\CC 2 01 $, and $\CC 2 34 $
with components 
$$   \Cc n uv st \equiv \intp \argpsq^{n-1} 
        f_{us}\argpsq f_{vt}\argpsq                        \eqnm bcndf $$
($\Cc n uv st = \Cc n vu ts $) exist and are invertible,
$$   \intp\, f_2\argpsq f_{0i}\argpsq = 0, \qquad
	\intp \psq\, f_2\argpsq f_{1i}\argpsq = 0 \,,      \eqnm orthb $$
$$   \OVER 14 \Cc 3/2 {\indfer\,\indfer} mn = \delta_{mn} \,,\quad
        \LF{ab} = -\OVER 12 \Cc 2 34 ab \,,  \quad
        \Lf{ij} = - \Cc 2 01 ij \,.                	   \eqnm Lmncd $$
If $\LF{ab}$ and $\Lf{ij}$ are given, relations \ref(Lmncd) are constraints on
$f_{3a}$, $f_{4a}$, $f_{0i}$, and $f_{1i}$; otherwise, relations \ref(Lmncd)
determine $\LF{ab}$ and $\Lf{ij}$.

To construct a tt extension of $\LNG$ we will use: scalar fields
$$   \phi_i[x;\sta] \equiv \Cc -1 00 ij \intp 
        f_{0j}\argpsq \sta_0\argxp \,, \quad     
     b[x;\sta] \equiv (\CC 1 22 )^{-1} \intp 
        f_2\argpsq \sta_0\argxp \,;                        \eqnm pbdef $$
chiral-bispinor fields
$$   \fsp_m[x;\sta] \equiv \Cc -1 {\indfer\indfer} mn
        \intp f_{\indfer n}\argpsq \pmatrix{ -\sigma_+\argp \cr I \cr} 
        \sta_\indfer\argxp \,;                             \eqnm fspmv $$
four-vector fields
$$   \eqalign{
     \varphi_i[x;\sta] &\equiv  2 \Cc -2 11 ij
        \intp f_{1j}\argpsq \sta_0\argxp \,p \,, \cr     
     A_a[x;\sta] &\equiv \OVER 43 \Cc -1 33 ab \intp 
        f_{3b}\argpsq (1 - \P_l) \sta_1\argxp \,;\cr}      \eqnm vAdef $$
and antisymmetric--rank-two--four-tensor fields
$$   F_a[x;\sta] \equiv \Cc -2 44 ab \intp f_{4b}\argpsq
        [ \sta_1\argxp \oot p - p \oot \sta_1\argxp ] \,.  \eqnm Fmdef $$
Here: $\CC -n uv $ denotes the matrix inverse to $\CC n uv $ (i.e.,  $ \Cc -n
uv st' \Cc n uv t't = \delta_{st} $); central, big brackets in \ref(fspmv) are
$4\times 2$ matrices; and $( a \oot b)^{\alpha\beta} \equiv a^\alpha b^\beta$
is the direct product of four-vectors $a$ and $b$. Fields
\ref(pbdef)--\ref(Fmdef) have properties \ref(macas) and \ref(macLt) required
of macroscopic variables; so,
$$   \eqalign{
        \phi_i[x;\stl] &= \phi_i\argx \,,       \cr 
        \fsp_m[x;\stl] &= \fsp_m\argx \,,       \cr 
        A_a[x;\stl] &= A_a\argx \,,             \cr}      \qquad 
     \eqalign{ 
        b[x;\stl] &= b\argx \,,                 \cr 
        \varphi_i[x;\stl] &= \varphi_i\argx \,, \cr
        F_a[x;\stl] &= F_a\argx \,.   \cr}      \qquad    \eqnm mvras $$ 
We will refer to the scalar and four-vector tt fields $\sta_0\argxp$ and
$\sta_1\argxp$ as the bosonic tt fields, and to the left-spinor tt field
$\sta_\indfer\argxp$ as the fermionic tt field, since they are related to boson
and fermion fields of $\LNG$ through \ref(pbdef)--\ref(mvras). 

Let us now consider the tt Lagrangian
$$   \Lext = \Lfree(\sta, \sd\sta) + \L_2(\sta) 
        + \L_3(\sta) + \L_\lambda(\sta) \,,                \eqnm ttLex $$
where: (1)~According to \ref(Lfdef), the free tt Lagrangian
$$   \displaylines{\quad
     \Lfree(\sta, \sd\sta) = \brck{\sta_0}{\sd\sta_0}{x} 
        + \half\brck{\sta_\indfer}{\sd \sta_\indfer}{x} \hfill\cr\hfill
        + \half\brck{\sd\sta_\indfer}{\sta_\indfer}{x} + 
	\brck{\sta_1}{\sd\sta_1}{x} \,.  \qquad\eqnmm(LfnAg) \cr} $$
For the asymptotic tt fields \ref(asnAs), $\Lfree$ equals the sum of all terms
in $\LNG$ that contain spacetime derivatives, by (A.18)--(A.20) and
\ref(Lmncd). (2)~We use
$$   \eqalignno{
     \L_2 &\equiv \LF{ab} F_{b\mu\nu}[x;\sta] \LF{ac} F_c^{\mu\nu}[x;\sta] 
        - \afsp_m[x;\sta] m_{mn} \fsp_n[x;\sta]      \cr
     &+ \half\Lf{ij} \varphi_{j\mu}[x;\sta] \Lf{ik} \varphi_k^\mu[x;\sta] 
        + \half\mu^2 \phi_i[x;\sta] \phi_i[x;\sta] 
	+ \mu^2 v^{-2} \Lb^2 b^2[x;\sta] \,,       \cr
     \L_3 &\equiv C_{acd} \LF{ab} F_b^{\mu\nu}[x;\sta] 
        A_{c\mu}[x;\sta] A_{d\nu}[x;\sta]    &\eqnmm(Lfrsp)\cr
     &- \afsp_m[x;\sta] \{ i\gamma^\mu A_{a\mu}[x;\sta] 
        T^a_{mn} + \phi_i[x;\sta] \Gamma^i_{mn} \} 
	\fsp_n[x;\sta]                   \cr
     &+ i \Lf{ij} \varphi_j^{\mu}[x;\sta] A_{a\mu}[x;\sta] 
        R^a_{ik} \phi_k[x;\sta] - \mu^2 v^{-2}\Lb b[x;\sta]
        \phi_i[x;\sta] \phi_i[x;\sta] \,.    \cr}                $$
Thus $\L_2(\stl)$ equals the sum of all quadratic terms in $\LNG$ that contain
no spacetime derivatives, whereas $\L_3(\stl)$ equals the sum of all cubic
interaction terms, by \ref(mvras). (3)~Real, scalar field
$$   \L_\lambda \equiv \lambda^{-2} \sum_j \brck{\sta_j}{(1 - \P_j) 
        \sta_j}{x}                                         \eqnm Llexp $$
with parameter $\lambda > 0$, $[\lambda ] = [\dol]$, and projections
$$   \eqalignno{
     \P_0\sta_0 &\equiv f_{0i}\argpsq \phi_i[x;\sta] + 2f_{1i}\argpsq 
        \varphi_i[x;\sta]\scpr p + f_2\argpsq b[x;\sta] \,,    \cr
     \P_\indfer \sta_\indfer &\equiv  \bigl( \sigma_-\argp , \I \bigr) 
        f_{\indfer m}\argpsq \fsp_m [x;\sta] \,,   &\eqnmm(Prdef)\cr
     \P_1\sta_1 &\equiv (1 - \P_l) f_{3a}\argpsq A_a [x;\sta] 
        + 2 f_{4a}\argpsq F_a[x;\sta] p                 \cr}      $$
[note that $\P_l\P_1 = \P_1\P_l = 0$, $\P_j\stlj = \stlj$, and
$\L_\lambda(\stl) = 0$]. The tt Lagrangian $\Lext$ defined by
\ref(ttLex)--\ref(Prdef) is a real, scalar field, by \ref(Lfdef) and since we
obtained its interaction part from the interaction part of $\LNG$ by replacing
its fields with the corresponding fields \ref(pbdef)--\ref(Fmdef) [which
transform under Lorentz transformations of the tt field $\sta$ the same way as
the original fields of $\LNG$, by \ref(macLt)], and then adding the real,
scalar field $\L_\lambda$.

If $\lambda \to 0$, either $\L_\lambda(\sta) = 0$, or $\L_\lambda(\sta)$ and
the tt Lagrangian $\Lext$ tend to infinity. Taking into account (A.7), 
$$   \brck{\P_j \sta_j}{\sta_j'}{x} = \brck{\sta_j}{\P_j \sta_j'}{x} \,,
        \qquad \P_j^2 = \P_j \,,  \quad j = 0, 1/2, 1,     \eqnm Pjprp $$
we can conclude that (1)~the Euler-Lagrange equations of $\L_\lambda(\sta)$
read
$$   (1 - \P_j ) \sta_j = 0 \,,   \qquad j = 0, 1/2, 1,    \eqnm PjEUL $$
and (2)~their most general solutions are the asymptotic tt fields \ref(asnAs).
As for every asymptotic tt field \ref(asnAs) the tt Lagrangian $\Lext$ equals
the QFT Lagrangian $\LNG$, i.e.,
$$   \Lext(\stl, \sd\stl) = \LNG \,,                        \eqnm Lrela $$
we may follow the arguments presented in Sec.~2.3 and consider $\LNG$ as a
path-integralwise approximation to $\Lext$ in the asymptote $\lambda \to 0$. So
we may presume that the tt Lagrangian $\Lext$ is a tt extension of the QFT
Lagrangian $\LNG$.

The gauge-invariant Lagrangian $\LC$ does not have mass terms needed to model
fundamental interactions. It is customary to get rid of this deficiency by
assuming that its gauge symmetry is spontaneously broken, and replace the
scalar fields $\phi_i\argx$ with $\phi_i\argx + \stgc\,$, $\stgc \stgc = v^2$.
This way $\LC$ acquires necessary, quadratic mass terms but no linear terms
[\cite[Cheng]]. If we accordingly shift the scalar fields in $\LNG$,
$$   \phi_i\argx \to \phi_i\argx + \stgc \,, \qquad
        b\argx \to b\argx + \half v^2/\Lb \,,              \eqnm trrep $$
then (1)~in $\LNG$ we do not generate linear terms excepting a divergence term,
and (2)~$\LNG$ becomes equivalent to $\LC$ whose scalar fields $\phi_i\argx$
have been replaced with the shifted fields $\phi_i\argx + \stgc$. The
$x$-independent tt field 
$$   \sta_{HG}\argp \equiv \Bigl( f_{0i}\argpsq \phi_{gi} +
        \half v^2 \ell^{-1} f_2\argpsq, 0, 0 \Bigr)\,,   
	\qquad \stgc\stgc = v^2 \,,                          \eqnm ttGHG $$
is a tt counterpart to the ground state $\phi_i\argx = \stgc$ and $b\argx =
\half v^2/\Lb$ of $\LNG$. Namely, the tt Lagrangian \ref(ttLex)--\ref(Prdef)
with $\sta \to \sta + \sta_{HG}$ has no linear terms (except for a divergence
term), and for $\sta = \stl$, it equals $\LNG$ whose scalar fields have been
shifted as specified by \ref(trrep).

\section Cubic interactions as the origin of quadratic ones

We now give an example of a tt Lagrangian, say, $\LCC$ that extends $\LNG$ as
specified in Sec.~2.3, and has only cubic interaction terms. Introducing real
functions $f\argpsq$ and $\fg\argpsq$ with dimensions $[\dol]^4$ and
$[\dol]^2$, and a positive dimensionless constant $a_g \equiv \brck{p f}{p
\fg}{x}$, we define:
$$  \LCC(\sta, \sd\sta) \equiv \Lfree + a_g^{-1} \brck{p f}{\sta_1}{x}\L_2
       + \L_3 + a_g^{-1} \brck{p f}{\sta_1}{x} \L_\lambda \,,
                                                           \eqnm LLGpr $$
where $\Lfree$, $\L_2$, $\L_3$, and $\L_\lambda$ are as defined in
\ref(LfnAg)--\ref(Prdef) with $\P_1$ replaced with $\P_1 + \P_l$. As the
variables of the path integral specified by $\LCC(\sta, \sd\sta)$ we take tt
fields $\sta$ of the form
$$   \sta = \sta_t + \stg \,,                              \eqnm stgdf $$
where $\sta_t \equiv (\sta_0, \sta_\indfer, (1 - \P_l)\sta_1 )$ and $\stg
\equiv (0, 0, p \fg)$; we note that
$$   \LCC\bigl( \sta_t + \stg, \sd(\sta_t + \stg) \bigr) = 
        \Lext(\sta_t, \sd\sta_t) \,,                       \eqnm LCsum $$
where $\Lext$ is the tt Lagrangian \ref(ttLex)--\ref(Prdef). If $\lambda \to
0$ and $\L_\lambda(\sta) \ne 0$, then $\LCC$ tends to infinity. The most
general solutions to the Euler-Lagrange equations of $\L_\lambda$ in
\ref(LLGpr) are $\stl + \stg$, which are allowed by \ref(stgdf). As $\LCC(\stl
+ \stg, \sd(\stl + \stg) ) = \LNG$, by \ref(LCsum), $\P_1\P_l = 0$, and
\ref(Lrela), we may indeed consider the tt Lagrangian $\LCC$ a tt extension of
the first-order Lagrangian $\LNG$.

If we replace $\sta$ in $\LCC(\sta, \sd\sta)$ with $\sta_t + \stg$, we generate
no linear terms, by \ref(LCsum). So we may see $\stg\argp$ as a ground state of
a kind, though there are certain distinctions between $\stg\argp$ and the tt
ground state $\sta_{HG}\argp$: (1)~Ground state $\stg$ is not determined by
the condition that the Lagrangian $\LCC$ gets no linear terms if we replace
$\sta$ with $\sta_t + \stg$. (2)~The ground state $\stg$ and the integration
variable $\sta_t$ in path integrals specified by $\LCC(\sta_t + \stg,
\sd(\sta_t + \stg) )$ do not interfere, by \ref(LCsum). Ground state $\stg$ is
an additional, $x$-independent, permanent, Lorentz-invariant feature of the
universe: a classical component of the relativistic vacuum extending throughout
the universe.

If we regard the tt Lagrangian $\LCC$ as the basic description of fundamental
interactions and the ground state $\stg\argp$ as a historical accident, then
$\LCC$ should not depend on constant $a_g$ determined by $\stg$ (for a given
$f$). This is the case if: (1)~We rescale the units of fields $A_a^\mu$ and
$\phi_i$ by replacing them in \ref(nAcdf) with $a_g^{-1/2} A_a^\mu$ and
$a_g^{-1/2} \phi_i$. (2)~We replace $\LF{ab}$ and $\Lf{ij}$ in \ref(nAlin) with
$a_g^{1/2} \LF{ab}$ and $a_g^{1/2} \Lf{ij}$. (3)~We assume that $C_{abc}$, $T^a
_{mn}$, $R^b_{ij}$, $\Gamma^i_{mn}$, $m_{mn}$, $\mu^2$, $v^2$, and $\lambda$
are proportional to $a_g^{1/2}$, $a_g^{1/2}$, $a_g^{1/2}$, $a_g^{1/2}$, $a_g$,
$a_g^2$, $a_g$, and $a_g^{-1/2}$, respectively. In such a case, $\lambda \to 0$
if $a_g \to \infty$, and we may explain the QFT asymptote $\lambda \to 0$ as
describing transitions where the interaction between the tt ground state $\stg$
and tt field $\sta_t$ suppresses projections $(1- \P_j)$ of its components so
much that they are negligible. Masses in $\LC$ with spontaneously broken
symmetry are then proportional to $a_g$. So we may interprete the mass terms in
$\LC$ with spontaneously broken symmetry as due to interaction with the
longitudinal, tt ground state $\stg$.

\section Unresolved questions about tt extensions

It is not clear which physical principles and properties, in addition to those
mentioned at the beginning of this section and to unitarity of the $S$-matrix,
further restrict realistic tt extensions of $\LNG$. The gauge invariance of
$\LC$ ensures renormalizability, also when it is spontaneously broken, and
severely constrains the values of $\LC$ parameters $C_{abc}$, $T^a_{mn}$,
$R^a_{ij}$, $\Gamma^i_{mn}$, and $m_{mn}$. We do not know whether there is a tt
counterpart to the concept of QFT gauge invariance (e.g., a symmetry and/or
high-energy unitarity constraints [\cite[Wein2]]) that similarly restricts
acceptable tt extensions of $\LNG$. Any restriction would be of great help
since there is an infinite variety of possible tt extensions of $\LNG$,
cf.~(A.22)--(A.31). When formulating a tt extension, we have, e.g., the
following options: 

(a)~The domain of the independent variable $p$. It may be the whole $\RRRR$ or
only a Lorentz-invariant subspace of $\RRRR$.

(b)~The kind of functions of $x$ and $p$ that tt fields $\sta\argxp$ are, 
cf.~Sec.~5.1 and [\cite[mi004]].

(c)~The number and type of tt fields $\sta_j\argxp$ used. We can use,
instead of the left-spinor tt field $\sta_\indfer\argxp$, a
right-spinor or a bispinor tt field: $\sta_{-\indfer}\argxp$ or
$\bigl(\sta_{-\indfer} \argxp, \sta_{\indfer}\argxp \bigr)$. We can
also use one scalar, one two-component-spinor, and one, possibly
transversal, four-vector tt field for each Higgs-scalar field
$\phi_i$, Dirac-spinor field $\fsp_m$, and four-vector gauge field
$A_a$, respectively. On the other hand, we can construct a tt
extension of $\LNG$ analogous to \ref(ttLex)--\ref(Prdef) without
using the scalar tt field $\sta_0$, by incorporating the fields
$\phi_i\argx$, $\varphi_i\argx$, and $b\argx$ of $\LNG$ into the
longitudinal component of the asymptotic, four-vector tt field
$\stlo$ instead of into the scalar $\stlz$ [e.g., by adding $\stlz p$
to $\stlo$ in \ref(asnAs)]. Alternatively, the tt Lagrangian
$\Lext(\sta_t, \sd \sta_t)$ in \ref(LCsum) is a tt extension of
$\LNG$ that needs fewer tt fields than $\Lext(\sta, \sd \sta)$,
namely $\sta_0$, $\sta_\indfer$, and $(1-\P_l)\sta_1$ instead of
$\sta_0$, $\sta_\indfer$, $(1-\P_l)\sta_1$, and $\P_l\sta_1$. If only
one four-vector tt field and one two-component-spinor tt field
suffice for modeling fundamental interactions, then no approximate
ttQFT can do with fewer tt fields, in contrast to QFTs where, e.g.,
QED uses fewer fields than the standard model.

(d)~The functions $f_{us}\argpsq$ that shape the $p$-dependence of
the asymptotic tt fields \ref(asnAs). Here they are restricted only
by the weak requirements \ref(bcndf)--\ref(Lmncd), but may not be
constant, and may also depend on $\lambda$. They are not necessarily
distinct, though those having the same first subscript are linearly
independent, by \ref(bcndf); e.g., we could choose $f_{3a}(y)$
proportional to $\sqrt y f_{4a}(y)$ and/or $f_{0i}(y)$ proportional
to $\sqrt y f_{1i}(y)$. We have yet to find out how to choose them so
that the fields \ref(pbdef)--\ref(Fmdef) are macroscopic variables.
Were the asymptotic tt fields $\stlj\argxp$ the initial parts of
certain expansions of tt fields $\sta_j \argxp$ in terms of
orthogonal functions of variable $p$, there would be additional
requirements on $f_{us}\argpsq$. Which might also imply restrictions
on the number and internal symmetries of QFT fields incorporated in
$\stlj\argxp$, cf.~(5.1), (5.3), and [\cite[mi006]].

(e)~The dependence on $\lambda$ and $\sta\argxp$. We can add to the
tt Lagrangians $\Lext$ and $\LCC$ any term that vanishes if $\lambda
\to 0$ or is finite as $\lambda \to 0$ and zero for $\sta\argxp =
\stl\argxp$.
 
(f)~A property of $\LNG$ is either explicitly displayed by its tt
extension, or it emerges only in the QFT asymptote. E.g., it is not
clear whether a tt Lagrangian that extends $\LNG$ should have a tt
counterpart to the gauge invariance and/or to the ground state of
$\LNG$, cf.~\ref(ttGHG).

If there is a basic ttQFT of fundamental interactions, better than
any QFT, it is of interest to know the answers to questions
raised in the above alternatives, and to the following ones:

(a)~How many free parameters has the basic tt Lagrangian in
comparison with the standard model, and how are they related?

(b)~Is the basic tt Lagrangian bilinear in fermionic tt fields?
cf.~[\cite[Wein3]].

(c)~Does the interaction part of the basic tt Lagrangian contain:
(1)~only cubic terms, due to the emission (creation) or absorption
(annihilation) of bosonic tt fields by fermionic or bosonic tt
fields; (2)~only quartic terms, due to the self-scattering of bosonic
tt fields or to the scattering of fermionic tt fields on bosonic
ones, but no quartic terms describing self-scattering of fermionic tt
fields; (3)~only cubic and quartic terms?  Comparing tt
Euler-Lagrange equations with the Boltzmann equation
[\cite[Libof]--\cite[Groot]], we would expect that the interaction
part of the basic tt Lagrangian contains only cubic terms. However,
noting that the Boltzmann equation models collisions of two
particles, we would expect only quartic terms.

(d)~Does a solution $\sta$ to the basic tt Euler-Lagrange equations
evolve toward an $x$-independent, Lorentz-invariant tt field, say,
$\steq$ as the time $t \to \infty$? If so: Is $\Pas\steq \ne 0$
and/or $(1 - \Pas)\steq \ne 0$? Is $\L(\steq, \sd\steq) = 0$? Is
$\steq$ related to a ground state that endows particles with mass?

(e)~How does it come about that we can adequately describe present
experiments by QFTs? What physical processes make it possible to
replace the basic ttQFT with some QFT? Can they be formally simulated
by limiting to zero some parameters of the basic ttQFT? Under what
conditions are $x$-independent terms absent from \ref(stjas) and
functions $\pfi k\argp$ independent of the initial and final states?

(f)~What are the implications for the QFT Lagrangian that
approximates the basic tt Lagrangian in the QFT asymptote? E.g., what
fields are present in the asymptotic tt fields $\stl\argxp$?
cf.~(A.8), and how are possible tt and QFT stationary points related?
cf.~\ref(ttGHG).

(g)~How are the quadratic terms with no spacetime derivatives
generated in Lagrangians of realistic QFTs? e.g., in $\LNG$ with
possibly shifted scalar fields \ref(trrep). In the kinetic theory of
gases, the linear terms of a tt Lagrangian would be determined by the
independent sources, whereas the bilinear ones would describe free
streaming or interactions with an underlying host medium not
described by tt fields. So it seems an appealing physical proposition
to assume, without any reference to possible symmetries of the basic
tt Lagrangian, that it has no linear terms, and no quadratic terms in
addition to those, \ref(Lfdef), that describe free streaming. If so,
there is a question about physical processes that result in a
realistic QFT Lagrangian with no linear terms and having also
quadratic terms with no derivatives. We see the following three
possibilities: First, quadratic terms with no derivatives in this QFT
Lagrangian could be generated within the QFT framework from a QFT
Lagrangian that has no such terms and is implied in the QFT asymptote
by the basic tt Lagrangian with no linear and quadratic-interaction
terms. Second, various non-exclusive tt processes could provide the
basic tt Lagrangian with counterparts to the quadratic terms with no
derivatives of the QFT Lagrangian in question: (1)~An
$x$-independent, Lorentz-invariant, tt ground state that (i)~does not
belong to the region of integration of ttQFT path integral,
(ii)~represents such a permanent classical relativistic vacuum as
$\stg$ does in \ref(LCsum), and (iii)~whose origin is not known.
(2)~A tt analogue to the QFT process of dynamical symmetry breaking
[\cite[Cheng]] that can be described in the low-energy approximation
by an effective ttQFT, the tt Lagrangian of which has the required
quadratic terms.  (3)~A tt phase transition analogous to the QFT
process of spontaneous symmetry breaking: An $x$-independent,
Lorentz-invariant stationary point of the basic tt Lagrangian is
added here to the original tt fields, which generates an equivalent
tt Lagrangian that has quadratic interaction terms but no linear
ones, cf.~\ref(ttGHG). And third, were the functions $\pfi{k}\argp$
in \ref(stjas), which shape the $p$-dependence of the asymptotic tt
fields $\stl\argxp$, actually certain averages of rapidly varying
functions of $x$, then we would have to take account of correlations
when evaluating the free tt Lagrangian \ref(Lfdef) for asymptotic tt
fields $\stl\argxp$. As a consequence, the resulting QFT Lagrangian
would acquire from the free tt Lagrangian also quadratic terms with
no derivatives. This case would be of special interest since mass
terms would appear only in the QFT asymptote.

\chapter Symmetry transformations

\section Spacetime translations

The tt Lagrangian \ref(Lgsum) is assumed to be invariant under spacetime
translations $x \to x + a$ in the sense of \ref(fLLin). [Translations $p \to p
+ a$ have no physical significance in tt approach; even the free tt Lagrangian
\ref(Lfdef) is not invariant under them.] So Noether's theorem for translations
implies that the energy-momentum tensor
$$   T^{\mu\nu}\argx \equiv \sum_j \Re \brck{\sta_j}{p^\mu \partial^\nu 
        \sta_j}{x} - \Lext(\sta,\sd\sta) \metrictensor^{\mu\nu} \,,
							   \eqnm fmten $$
where $\Re$ stands for real part of, satisfies the continuity equation
$$   \partial_\mu T^{\mu\nu}\argx = 0			   \eqnm coneq $$
provided tt fields $\sta_j\argxp$ satisfy the tt Euler-Lagrange equations
[\cite[Wein4]\cite[Sudbe]]. The energy-momentum tensor of the tt ground state
$\stg\argp$ is a relativistic invariant $T^{\mu\nu}\argx \equiv 0$, by
\ref(LCsum).

\section Internal symmetries

We will say that a tt Lagrangian $\Lext(\sta,\sd\sta)$ is invariant under
infinitesimal, internal-symmetry transformations 
$$   \sta_j\argxp \to \sta_j(x,p;\epsilon) \equiv \sta_j\argxp +
        i\epsilon \sum_k t_{jk} \sta_k\argxp 		   \eqnm intsy $$
if
$$   {\partial\over \partial\epsilon} \Lext \bigl( \sta(x,p;\epsilon),
         \sd\sta(x,p;\epsilon) \bigr) \Big|_{\epsilon=0} = 0 \,,
                                                           \eqnm ints3 $$
where $\epsilon$ is a parameter and $t_{jk}$ a mapping of the tt field
$\sta_{k}$ into the tt field $\sta_j$. If so, the associated current
$$   j\argx \equiv \sum_{jk} \Re \brck{\sta_j}{ip\, t_{jk} \sta_k}{x}  
                                                           \eqnm intcu $$
satisfies the continuity equation
$$   \partial\scpr j\argx = 0                              \eqnm cont2 $$
provided $\sta_j$ satisfy the tt Euler-Lagrange equations, by \ref(Lgsum) and
\ref(Lfdef) [\cite[Wein4]\cite[Cheng]\cite[Sudbe]]. Infinitesimal
transformations \ref(intsy) appear the same in all inertial frames if, and only
if $U(\Lambda, a) t_{jk} = t_{jk} U(\Lambda, a)$; in such a case, the current
$j\argx$ transforms as a four-vector field under the replacement of $\sta
\argxp$ with $U(\Lambda, a)\sta\argxp$. If $t_{jk}$ are such complex constants
that $t_{kj}^* = t_{jk}$, then the current \ref(intcu) is invariant under
infinitesimal transformations \ref(intsy), i.e., $\partial j(x) \big/ \partial
\epsilon = 0 $ at $\epsilon = 0$.

For a global phase change of the tt field $\sta_j\argxp$,
$$   \sta_j\argxp \to e^{i\varphi} \sta_j\argxp \,,
          \qquad \varphi \in \RR \,,			   \eqnm phsym $$
the associated current $j\argx = \brck{\sta_j}{ip \sta_j}{x}$ is a real,
four-vector field, by (A.4) and (A.6). [This current vanishes for real $\sta_j
\argxp$, $j=0$, $1$, by (A.4)--(A.6): a real bosonic tt field effects no such
current.] Both tt Lagrangians \ref(ttLex)--\ref(Prdef) and
\ref(LLGpr)--\ref(stgdf) exhibit invariance under global phase changes of the
fermionic tt field $\sta_\indfer \argxp $, which is analogous to the invariance
of Lagrangians $\LC$ and $\LNG$ under joint global phase changes of fermion
fields $\fsp_m$.

Lagrangian $\LNG$ is invariant under the infinitesimal, global, gauge
transformations
$$   \eqalign{
     A_a \to A_a + \epsilon_b C_{abc} A_c \,,    &\qquad
        \LF{ab} F_b \to \LF{ab} F_b + \epsilon_b C_{abc} 
	   \LF{cd} F_d \,,                       \cr
     \fsp_m \to \fsp_m - i\epsilon_a T^a_{mn} \fsp_n \,,   &\qquad
        \Lb b \to \Lb b \,,                      \cr
     \phi_j \to \phi_j - i\epsilon_a R^a_{jk} \phi_k \,,   &\qquad
     \Lf{ij} \varphi_j \to \Lf{ij} \varphi_j - i\epsilon_a R^a_{jk} 
        \Lf{kl} \varphi_l \,,                    \cr}      \eqnm ggtra $$
where $\epsilon_a$ are real parameters [\cite[Cheng]]. As the tt extension
\ref(ttLex)--\ref(Prdef) of $\LNG$ does not exhibit an analogous symmetry, let
us point out such a tt extension of $\LNG$ that is invariant under
infinitesimal, global--internal-symmetry transformations analogous to
\ref(ggtra). To this end, we take the following tt fields: real, scalar tt
fields $\sta_0 \argxp$ and $\sta_{0j}\argxp$; complex, left-spinor tt fields
$\sta_{\indfer m} \argxp$; real, four-vector tt fields $\sta_{1a}\argxp$;
and associated asymptotic tt fields
$$   \eqalign{
     \sta_{0\as} &= f_0\argpsq b\argx \,, \qquad
     \sta_{0j\as} = f_0\argpsq \phi_j\argx +
        2 f_1\argpsq p \scpr \varphi_j\argx \,,    \cr
     \sta_{\indfer m\as} &= f_\indfer\argpsq \bigl( \sigma_-\argp, I \bigr)
        \fsp_m\argx \,,                          \cr
     \sta_{1a\as} &= f_2\argpsq A_a\argx +
        2 f_3\argpsq F_a\argx p \,.              \cr}      \eqnm ginas $$
We then construct a tt extension of $\LNG$ on the analogy of the tt extension
\ref(ttLex)--\ref(Prdef). Such tt extension of $\LNG$ is invariant under
infinitesimal, global--internal-symmetry transformations
$$   \eqalign{
     \sta_{1a} &\to \sta_{1a} + \epsilon_b C_{abc} \sta_{1c} \,,     \cr
     \sta_{\indfer m} &\to \sta_{\indfer m} 
        - i\epsilon_a T^a_{mn} \sta_{\indfer n} \,,        \cr
     \sta_{0j} &\to \sta_{0j} - i\epsilon_a R^a_{jk} \sta_{0k}\,, 
        \qquad \sta_0 \to \sta_0 \,;     \cr}     \eqnm ttgtr $$
and the associated conserved currents \ref(intcu) are $j_a\argx \equiv - \Re
\brck{ \sta_{\indfer m}}{ ip T^a_{mn} \sta_{\indfer n}}{x}$. Hence, there are
tt extensions of $\LNG$ that exhibit an analog to the global--non-Abelian-gauge
invariance, though they need more tt fields than tt Lagrangians
\ref(ttLex)--\ref(Prdef) and \ref(LLGpr)--\ref(stgdf), which exhibit less
symmetry [see (A.28)--(A.34) about the local gauge invariance]. So in ttQFTs,
more symmetry requires more tt fields $\sta_j$, since the free tt Lagrangian
\ref(Lfdef) is not invariant under operations on macroscopic variables of its
tt fields $\sta_j$; a simpler tt model exhibits less symmetry! This makes one
wonder whether ttQFTs of fundamental interactions exhibit global or local,
non-Abelian--gauge symmetries only in the QFT approximation [\cite[Frogg]];
especially as one has to add terms to Lagrangians of QFTs to remove the gauge
invariance, thereby allowing construction of propagators.

\section $\C$, $\P$, $\T$, and chiral transformations 

For scalar, four-vector, left-spinor, and right-spinor tt fields we define the
{\it charge conjugation transformation} $\C$ as follows:
$$  \eqalign{
    \C\sta_0 &\equiv \sta_0^*(x,p)\,,          \qquad
       \C\sta_1 \equiv -\sta_1^*(x, p) \,,     \cr
    \C\sta_{\pm\indfer} &\equiv \mp\sigma_\mp\argp \sigma^2 
       \sta_{\pm\indfer}^*(x,p) \,.           \cr}      \eqnm charg $$
Note that $U(\Lambda, a) \C = \C U(\Lambda, a)$, $\C^2 = 1$, and $\brck{ \C
\sta_j}{ \C \sta_j'}{x} = (-1)^{2j} \brck{\sta_j}{ \sta_j'}{x}^*$. Hence,
$\brck{ \sta_j}{\sta_j}{x} = \OVER 14 \brck{(1 + \C)\sta_j}{ (1 + \C)\sta_j}{x}
+ (-1)^{2j}\, \OVER 14 \brck{(1 - \C)\sta_j}{ (1 - \C)\sta_j}{x}$, i.e.,
$(1+\C)\sta_j$ and $(1-\C)\sta_j$ do not interfere. In particular, real and
imaginary parts of $\sta_0$ and of $\sta_1$ do not interfere.

For tt fields we define the {\it space inversion transformation} $\P$ as
follows:
$$  \eqalign{
    \P\sta_0 &\equiv \sta_0(\P_1 x,\P_1 p) \,, \qquad
    \P\sta_1 \equiv (\P_1\sta_1)(\P_1 x,\P_1 p)\,,        \cr
    \P\sta_{\pm\indfer} &\equiv \sigma_\mp\argp 
       \sta_{\pm\indfer}(\P_1 x, \P_1 p)\,,        \cr}      \eqnm parit $$
where $\P_1 x \equiv (x^0, -\vx)$ is the space inversion of a four-vector $x$.
Note that $\P^2 = 1$, $\brck{\P \sta_j}{ \P \sta_j'}{x} = \P \brck{\sta_j}{
\sta'_j}{x}$, and $U\bigl(\Lambda(\gg,\bb), a \bigr) \P = \P U\bigl(\Lambda(
\gg, -\bb), \P_1a \bigr)$ with $\gg$ and $\bb$ being the rotation and boost
parameters of a proper ortochronous Lorentz transformation $\Lambda$.

For a four-vector $x$, its time-inversion $\T_1 (x^0, \vx) \equiv (-x^0, \vx)$.
For tt fields we define the {\it time-reversal transformation} $\T$ as follows:
$$  \eqalign{
    \T\sta_0 &\equiv \sta_0^*(\T_1 x,\T_1 p)\,,  \qquad
    \T\sta_1 \equiv - (\T_1\sta_1^*)(\T_1 x, \T_1 p) \,, \cr
    \T\sta_{\pm\indfer} &\equiv -i\sigma^2\sta_{\pm\indfer}^*
       (\T_1 x, \T_1 p) \,.                      \cr}	   \eqnm timeS $$
Note that $U\bigl(\Lambda(\gg,\bb), a \bigr) \T = \T U\bigl( \Lambda(\gg,-\bb),
\T_1 a \bigr)$, $\T^2\sta_j = (-1)^{2j} \sta_j$, and\break $\brck{ \T \sta_j}{
\T \sta_j'}{x} = \T \brck{\sta_j}{\sta'_j}{x}$.

We may regard the above tt symmetry transformations $\C$, $\P$, and $\T$ as
extensions of the QFT symmetry transformations $\C$, $\P$, and $\T$ with
certain phase factors, because under $\C$, $\P$, and $\T$ transformations of tt
fields: (1)~the asymptotic fields \ref(asnAs) satisfy relations analogous to
\ref(stlco), and (2)~the fields \ref(pbdef)--\ref(Fmdef) transform as the
corresponding QFT fields; e.g., $\xfi{m}[x; \C\sta] = \gamma^2 \xfi{m}^* [x;
\sta]$, $\xfi{m}[x; \P\sta] = - \gamma^0 \xfi{m} [\P_1 x; \sta]$, and $\xfi{m}
[x;\T\sta] = \gamma^3 \gamma^1 \xfi{m}^* [\T_1 x; \sta]$, by \ref(fspmv) and
\ref(charg)--\ref(timeS). [This is so only when the functions 
$f_{us} \argpsq$ in \ref(asnAs) are real-valued.] If a scalar or a four-vector
tt field, $\sta_0 \argxp$ or $\sta_1 \argxp$, does not depend on the additional
variable $p$, then the tt symmetry transformations $\P$, $\T$, and $\C$ have
the same effect as the original QFT ones. 

Any product of tt transformations $\C$, $\P$, and $\T$ is equivalent to the
strong-reflection tt transformation; i.e., for $j = 0$, $\pm 1/2$, $1$, we have
$$  \displaylines{\quad
    \C\P\T\sta_j = (-1)^{2j}\P\C\T\sta_j = \P\T\C\sta_j = \T\P\C\sta_j =
       (-1)^{2j}\T\C\P\sta_j \hfill\cr\hfill{}= 
    \C\T\P\sta_j = i^{-2j} \sta_j(-x,-p) \,. 
                                   \qquad\hbox{\eqnmm(strrf)}\cr}  $$
Under the combined $\C$, $\P$, $\T$, and $x \to -x$ transformations of tt
fields $\sta_j$, taken in any order, the terms of the tt Lagrangian
\ref(ttLex)--\ref(Prdef) that depend bilinearly on the fermionic tt field
$\sta_\indfer\argxp$ change sign, whereas the terms depending only on the
bosonic tt fields $\sta_0 \argxp$ and $\sta_1 \argxp$ remain the same---just as
in the case of $\LNG$ or $\LC$. On the analogy of the CPT theorem of QFTs, we
expect that also the theoretical results of realistic ttQFTs are invariant
under combined $\C$, $\P$, and $\T$ transformations, i.e., {\it we expect that
fundamental interactions exhibit the CPT invariance also in ttQFTs.}

We define the {\it chiral transformation} of spinor tt fields as
$$	\sta_{\pm\indfer}\argxp\to
		\mp \sta_{\pm\indfer}(x, -p) \,,	   \eqnm chirt $$
It induces the QFT chiral transformation $\xfi m [x; \sta_{\pm\indfer} ] \to
\gamma^5 \xfi m [x; \sta_{\pm\indfer} ]$ in the asymptotic tt fields
\ref(asnAs) and (A.8), and in the chiral-bispinor fields \ref(fspmv) and
(A.13). Free tt Lagrangian \ref(Lfdef) is invariant under tt chiral
transformation \ref(chirt).

\chapter Comments on physics of ttQFTs

In the tt approach to quantum dynamics, the first-order form $\LNG$ of a
general gauge-invariant Lagrangian represents physical processes in the QFT
asymptote, though it is equivalent to the second-order form $\LC$. So due to
the form \ref(Lfdef) of the free tt Lagrangian, the tt approach suggests that
the first-order formalism of QFTs is a more direct description of fundamental
interactions than the conventional, second-order one. Which (1)~may explain
claims that the first-order formalism simplifies calculations in QFTs
[\cite[McKeo]], (2)~concurs with Schwinger's statement that it is ``natural''
[\cite[McKeo]], and (3)~gives support to Greiner's conjecture that ``the Good
Lord wrote the field equations in linearized form'' [\cite[Grein]], since the
Euler-Lagrange equations of a first-order-form QFT Lagrangian are themselves
first-order partial-differential equations.

According to ttQFTs, all spacetime derivatives in Lagrangians of QFTs,
including those that appear in derivative couplings, come from the free part
\ref(Lfdef) of tt Lagrangians. Thus the tt approach offers a simple, unified
physical explanation of the origin of all spacetime derivatives in QFT
Lagrangians: the free streaming of Feynman's X-ons in between strictly local
collisions, which is described by the substantial time derivative $\sd$.

In QFTs one can derive the appropriate path-integral formalism from the
canonical formalism, and vice versa [\cite[Wein4]]. But within the tt framework
this may not be so outside the QFT asymptote. If nature is nonlocal or granular
in the small, Bjorken and Drell [\cite[Bjork]] expect that the canonical
formalism applies only in the sense of a correspondence principle for large
distances.

As in the QFT asymptote $\lambda$ is presumed to be, in effect, infinitesimal,
we expect the physics beyond QFTs to be characterized also by a very large
energy scale $\lambda^{-1}$. As the Planck length is absent from present QFTs,
vastly smaller than any characteristic length of processes described by them,
and gives the distance and energy at which present QFTs are expected to break
down, it is possible that $\lambda \equiv \sqrt{ \hbar G/c^3}$. If so, the
Planck length is a fundamental unit of length in ttQFTs. However, it is not
clear how to take account of gravitational forces within the tt framework.
E.g., we do not know: (1)~Is gravity inherent to tt approach? since there are
such local averages of bosonic tt fields $\sta_j \argxp$, $j =0$, $1$, that
transform under Lorentz transformations of $\sta_j\argxp$ as symmetric-tensor
fields [e.g., (A.15)]. (2)~Should the asymptotic tt fields contain the
gravitational, symmetric-tensor field? i.e., can gravity be modeled by a QFT or
does it qualitatively modify QFTs? cf.~[\cite[Feyng]], (A.8), (A.18), and
(A.20).

Certain theoretic considerations imply the existence of {\it causal,
superluminal influences} that cannot be directly observed, though
they are an essential component of quantum phenomena. Without them it
is hard to explain, e.g., experimental violations of Bell's
inequalities [\cite[mi001]] and recent experiments in quantum optics
[\cite[Chiao]], which imply that distant events can causally
influence each other faster than any light signal could have
travelled between them. Present QFTs give adequate, probabilistic
descriptions of these experiments; yet they give no clue about the
physical processes that could effect such causal, superluminal
influences. But ttQFTs may give some, since their Euler-Lagrange
equations may be regarded as describing the classical transport of
Feynman's X-ons whose ``speeds'' $c|\vp/p^0|$ are not bounded. In the
QFT asymptote, the additional variable $p \in \RRRR$ is averaged out,
but we expect the physics beyond QFTs to be characterized also by
these infinite ``speeds''. We were able to give examples of tt
Euler-Lagrange equations that exhibit causal, superluminal
influences, even though they do not propagate changes in their
macroscopic variables away from their sources faster than light
[\cite[mi001]].

Physical significance of Feynman's X-ons is open to discussion. The
concept of pointlike entities whose four-momenta $p$ comprise the
whole $\RRRR$ opens a whole new vista in quantum metaphysics
[\cite[Sudbe]] that we did not consider.  Pursuing Feynman's
atomistic idea, we pretended that the partial-differential equations
of QFTs model the macroscopic motion of hypothetical X-ons. On the
analogy of the kinetic theory of gases, we then surmised that the
evolution of such macroscopic motion can be better described by
integro-differential, transport equations that asymptotically imply
partial-differential equations of QFTs [\cite[mi004]]. Hence, as an
exploratory step we proposed that QFT Lagrangians be replaced by
appropriate tt Lagrangians, and Feynman's path integrals defined
accordingly [\cite[mi002]].  Which is the whole significance of X-ons
in this paper; here, we may interprete the independent variable $p$
just as an additional, continuous index of ``fields'' $\sta_j\argxp$.
In this connection, the following Polyakov's remarks [\cite[Polya]]
are worth noticing: ``Elementary particles existing in
nature resemble very much excitations of some complicated medium
(ether). We do not know the detailed structure of the ether but we
have learned a lot about effective Lagrangians for its low energy
excitations. It is as if we knew nothing about the molecular
structure of some liquid but did know the Navier-Stokes equation and
could thus predict many exciting things. Clearly, there are lots of
different possibilities at the molecular level leading to the same
low energy picture.''

\section QFTs with an infinite number of fields

On the analogy with the polynomial expansions and the method of
moments in particle transport theory [\cite[Libof]--\cite[Groot]], we
expect that any tt field $\sta\argxp$ can be completely described by
an infinite number of relativistic fields that are certain averages
over the variable $p \in \RRRR$, cf.~\ref(Stjas). Hence, in tt
approach, we expect the Lagrangians of QFTs modeling more accurate or
higher-energy experiments to be customized and contain not only more,
specific terms, but also appropriate additional fields.  Which is in
line with our experiences with the standard model. For instance,
considering measured values of the muon anomalous magnetic moment,
Kinoshita and Marciano [\cite[Kinos]] pointed out that pure QED
suffices to explain only the first four significant digits, using to
this end 16 field-components altogether; whereas to obtain a complete
explanation of the first five significant digits, one needs QED and
QCD with additional 104 field-components.  Moreover, Brodsky
[\cite[Brods]] pointed out that at higher energies beyond standard
QCD, it is plausible that new fields with higher color representation
will be needed, and it is conceivable that the present quark and
gauge fields are themselves composite at short distances. Well, as we
pointed out above, infinitely many fields are actually needed for a
complete modeling of fundamental interactions in tt approach; but
their physical significance is not clear. So we do not know: how they
manifest themselves, what kind of matter or forces they represent,
how many of them represent physical particles, how many fields of the
standard model are among them, whether the asymptotic tt fields
contain only a finite number of them, whether they are useful only in
certain combinations, and what kind of constraints they are subject
to, cf.~photons, gauge fields, Faddeev-Popov ghost fields, and
quarks.

Let us consider a ttQFT whose tt fields are sums
$$   \sts\argxp \equiv \sum_l \pfi l\argp \xfi{l}\argx     \eqnm infsm $$
of appropriate products of certain given functions $\pfi{l}\argp$ of $p \in
\RRRR$ and of relativistic fields $\xfi{l}\argx$ that specify the
$x$-dependence of a particular tt field $\sts\argxp$. If we require that the
substantial derivative $\sd$ of such a sum is also a sum of the same kind,
these sums cannot be finite as the presence of certain $\pfi l\argp$ implies
the presence of $\pfi l\argp p^\mu$, $\pfi l \argp p^\mu p^\nu$, $\ldots$;
which suggests how to construct the smallest set of tt fields $\sts\argxp$ that
contains given asymptotic tt fields. On the analogy with the asymptotic tt
fields $\stl\argxp$ and their macroscopic variables, we assume: (1)~There are
such mappings $\xfi{l}[x; \sts]$ of tt fields $\sts\argxp$ into relativistic
fields that 
$$   \xfi{l}[x; \sts] = \xfi{l}\argx \,.                   \eqnm xfone $$
(2)~For every inhomogeneous Lorentz transformation we have
$$   U(\Lambda, a) \sts\argxp =
        \sum_l \pfi{l}\argp U(\Lambda, a) \xfi{l}\argx \,, \eqnm xfLor $$
so that $\xfi{l}[x; U(\Lambda,a) \sts] = U(\Lambda,a) \xfi{l}[x;
\sts]$. So the tt Lagrangian $\Lext(\sts, \sd \sts)$ of this ttQFT
depends solely on an infinite number of relativistic fields $\xfi{l}
\argx$ and their first-order derivatives $\partial \xfi{l}$, say,
$\Lext(\sts, \sd \sts) = \Linf( \ldots, \xfi{l}, \partial \xfi{l},
\ldots)$, and transforms as a scalar field under Lorentz
transformations $\xfi l \to U(\Lambda,a)\xfi{l}$. Thus the ttQFT in
question is actually a QFT with an infinite number of fields $\xfi{l}
\argx$. The main characteristic of such a QFT is that its Lagrangian
$\Linf$ contains derivatives $\partial \xfi{l}$ of its fields
$\xfi{l}\argx$ only linearly and solely in its standard free part
\ref(Lfdef), which couples them bilinearly with various fields
$\xfi{l}$, cf.~(A.18)--(A.20). On the analogy of \ref(Lttft), we may
regard the Lagrangian $\Linf$ as a tt extension of a certain QFT
Lagrangian $\Lq$ if $\Linf$ equals $\Lq$ when we put all but a finite
number of $\xfi{l}$ equal to zero. If the fields $\xfi l\argx$ of
$\Linf$ that are common to $\Linf$ and $\Lq$ are also the macroscopic
variables of tt field $\sts\argxp$, they are always constrained by
the initial and final states just like the same fields of $\Lq$. How
the values of the remaining fields $\xfi l\argx$ are constrained at
the initial and final instants is an open question. It is possible
that at least some of them do not depend on the initial and final
states, e.g., they might be free or equal to zero.

\section Transition from a ttQFT to the QFT it extends

In the kinetic theory of gases, physical processes described by the
Boltzmann transport equations exhibit asymmetry with respect to the
direction of time [\cite[Libof]\cite[Grad]]. A perturbation of a
classical gas in equilibrium presumably disappears in three stages
with widely disparate time scales: (1)~A microscopic phase, during
which the $p$-dependence of tt field rapidly simplifies as tt field
tends toward an asymptotic tt field with characteristic
$p$-dependence, and a smooth, slowly varying $x$-dependence. The
dynamics of microscopic phase is as yet beyond direct laboratory
measurements. (2)~A subsequent macroscopic phase, which is
characterized by smooth and slowly varying macroscopic variables.
These fields, whose evolution can now be adequately modeled by the
partial-differential equations of fluid dynamics, determine the
$x$-dependence of the asymptotic tt field. (3)~As time $t \to
\infty$, tt field very slowly approaches an $x$-independent,
Lorentz-invariant, equilibrium tt field.  The preceding applies also
to a perturbation of a macroscopic phase. This classical behaviour
makes us propose the {\it hypothesis} that the quantum dynamics of
fundamental interactions exhibits an analogous temporal behaviour
with an ``arrow of time'' in the following sense: For a given initial
state at $ t=0$, there is (i)~some QFT that is extended by the basic
ttQFT of fundamental interactions, and (ii)~a Lorentz-invariant
subset of states, say, the asymptotic states such that: (1)~The ttQFT
probability for a transition from the initial state to any final
state at instant $\tfin$ but an asymptotic one is negligible when
$\tfin$ is large enough, i.e., after a while, the ttQFT probability
for finding by measurements any non-asymptotic state becomes
negligible. (2)~The ttQFT probability for a transition from an
initial asymptotic state to any non-asymptotic state is negligible.
So physical processes are practically irreversible in the sense that
non-asymptotic initial states eventually result in asymptotic final
states, but the reverse is not to be expected. (3)~As far as the
present-day experiments are concerned, the ttQFT probability
amplitude for a transition between two asymptotic states is
adequately approximated by the QFT one. Thus points (1)--(3) offer an
explanation how phenomena that we can model by QFTs eventually
emerge. (4)~When QFT is applicable and transition amplitude is not
negligible, the predominant contribution to the QFT path integral
comes from such paths that do not vary appreciably over time and
space intervals comparable to $\lambda$. (5)~There is a particular,
$x$-independent final state, the equilibrium state, such that the
ttQFT probability for transition from the initial state to any final
state at instant $\tfin$ but to the equilibrium state becomes
negligible as $\tfin \to \infty$. Thus, as $t \to
\infty$ all radiation and matter disappear.

Our physical picture of how transition amplitudes evolve is entirely
based on our experience and insights gained from such quantum
processes where QFTs apply, which use a finite number of fields. So
we cannot expect this picture to be of much help in comprehending the
evolution of transition amplitudes where only some ttQFT can provide
an adequate description, using in fact an infinite number of fields.
However, it seems reasonable to assume that the probability
amplitudes for transitions between initial and final states of
several particles can be modeled by some QFT so long as these
particles stay sufficiently apart in the meantime. We think that such
initial, ``in'' states of scattering experiments in nuclear or
elementary particle physics that result in ultrarelativistic
multiparticle interactions do not belong to the asymptotic states
(though the final, ``out'' states do). If so, we might obtain
information about the hypothetical tt properties of fundamental
interactions by considering in detail such multiparticle
interactions.

If one extrapolates backward various relativistic cosmological
models, they imply under quite general conditions that sometime in
the past the whole content of the universe was located within a small
region, which then expanded outward. On the analogy of the transition
from transport, integro-differential equations to the hydrodynamic,
partial-differential equations, we therefore expect those concepts
(particles and forces; fields; and dynamic laws with their constants,
symmetries, and conservation laws) that are used to model the present
universe to be of limited use for modeling the very early universe
(or for that matter a white dwarf, a neutron star, or a black hole).
The very early, in variable $x$ extremely localized universe was
probably dominated by the microscopic phenomena characterized by
unlimited ``speeds'' $c |\vp/p^0|$, and its initial expansion faster
than the subsequent one dominated by macroscopic movements whose
speeds are limited by the speed of light.

A transition from a ttQFT to the QFT it extends is mathematically
highly singular since it involves transition from an infinite number
of fields to a finite number of them. So the results of a ttQFT have
in general an essential singularity at $\lambda = 0$, i.e., at most
an asymptotic expansion in powers of $\lambda$. And a QFT result is
in general not the first term of a convergent expansion in powers of
$\lambda$ of the corresponding ttQFT result.

\chapter Concluding remarks

In this paper we have put forward a new class of relativistic,
quantum field theories in $\RRRR \times \RRRR$, ttQFTs, which may
extend the present QFTs to higher energies without neccesarily
implying any new particles whatsoever. They are based on the Feynman
path-integral approach to quantum dynamics, without changing in any
way the basic conceptual framework of QFTs; in particular, the
spacetime $\RRRR$ and quantum statics [\cite[Sudbe]] remain
unchanged. We have patterned the proposed ttQFTs on the kinetic
theory of rarefied gases [\cite[Libof]\cite[Groot]\cite[Grad]], which
gives a {\it more detailed} and {\it accurate} description of
physical processes than fluid dynamics, which is suitable only for
modeling large-scale phenomena. Therefore, we hope the proposed tt
approach to quantum dynamics will also enable us to find out more
about the small-scale physics of fundamental interactions that
underlies phenomena so far described by QFTs. This tt approach is, to
our knowledge, the first-ever attempt to apply Feynman's idea about
granularity of the microscopic world [\cite[Feynm]] to quantum
dynamics of fundamental interactions. His idea is an expression of
the ancient atomistic concept, well proven in physics of fluids and
solids [\cite[Wein1]], which we took into account in a continuous
manner on the analogy with the Boltzmann transport equation.

In Section~2, we specified the following basic assumptions of an
extension of a QFT by a ttQFT: (1)~Quantum processes considered can
be described by the initial and final states of the QFT. (2)~When
calculating transition amplitudes, we replace the QFT path integrals,
whose paths are fields of the spacetime variable $x$, with 
path integrals whose paths are tt fields of two independent, real,
four-vector variables $x$ and $p$. (3)~Lagrangian of the ttQFT is a
sum of a free Lagrangian, whose Euler-Lagrange equations are the
partial-differential equations of free streaming, and of an
interaction Lagrangian that is local in $x$-variable and contains no
derivatives. (4)~The interaction ttQFT Lagrangian depends on a
positive parameter $\lambda$. In general, it tends to infinity if
$\lambda \to 0$, but not for the asymptotic tt fields that are:
(i)~the most general solutions to the ttQFT Euler-Lagrange equations
up to the leading order of $\lambda$, and (ii)~sums of QFT fields
multiplied by certain functions of variable $p$. (5)~The QFT
description of a quantum process can be obtained from the extending
ttQFT in the asymptote $\lambda \to 0$, where ttQFT transition
amplitudes equal the QFT ones. As the ttQFT extending the QFT implies
it when $\lambda$ becomes exceedingly small, ttQFT may be regarded as
a generalization in the sense of Weinberg [\cite[Wein3]]. (6)~In the
asymptote $\lambda \to 0$, the asymptotic tt fields whose QFT fields
are consistent with the initial and final states may be taken as the
domains of integration of ttQFT path integrals. (7)~Covariant
averages of tt fields exist that have the same physical significance
as the QFT fields.

In Section~3, we constructed a tt Lagrangian whose path integrals are
formally equivalent in the asymptote $\lambda \to 0$ to the path
integrals specified by a general, first-order, gauge-invariant
Lagrangian. Which demonstrates that extensions of QFTs are feasible
within the tt framework; and we can regard the standard model as an
asymptotic approximation analogous to the fluid dynamics in the
kinetic theory of gases. We do not know how many tt fields are
actually needed for modeling fundamental interactions. Only two may
be needed, one for fermions and one for bosons. If so, the different
fermion, bispinor fields of the standard model are only different
manifestations, covariant averages, of the same two-component--spinor
tt field. As there are infinitely many possible tt extensions of a
QFT Lagrangian, in Sec.~3.3 we pointed out certain available options.
A crucial unresolved question is what physical principles
characterize physically relevant tt extensions, and, possibly, even
uniquely determine them up to a few adjustable parameters.

In Section~4, we considered tt symmetry transformations that induce
in asymptotic tt fields the QFT transformations of QFT fields.
However, it is not clear which symmetries of a QFT Lagrangian should
be required of an extending tt Lagrangian: tt symmetry principles
obeyed by fundamental interactions have yet to be identified.

In Section~5, we touched on physics beyond QFTs that the tt approach
suggests on the analogy of the kinetic theory of gases. We pointed
out such ttQFTs that are actually QFTs with (i)~an infinite number of
fields, (ii)~massless, first-order, free Lagrangians, and (iii)~no
derivative couplings. We discussed a transition from a ttQFT to a
QFT.

In this paper, we have presented the basic premises of a new, tt
approach to the extension of QFTs to higher energies, and indicated
possibilities and open questions. One has yet to determine realistic,
internally consistent ttQFTs of fundamental interactions, find out
their underlying principles and physical contents (e.g., how they
describe classical systems and infered superluminal influences), and
work out typical, observable, qualitative and quantitative
consequences (in particular, experimentally testable corrections to
QFTs). Preliminary identification and study of the simplest,
relevant, model ttQFTs is probably necessary to gain an understanding
of realistic ttQFTs: how they work, how they modify dynamic concepts
of QFTs, what kind of physics beyond QFTs they can describe, how
rigid and logically isolated they are in the sense of Weinberg
[\cite[Wein1]], what approximations and mathematical methods make
sense and when, and what methods and results of contemporary QFTs in
higher dimensions are applicable to ttQFTs.

\bigbreak\leftline{\chapterfont Acknowledgements}
\nobreak\medskip
We thank M.~Polj\v sak and A.~Verbov\v sek for many useful
discussions.

\appendix 

In this Appendix we have collected relations and examples of interest
in constructing tt extensions of Lagrangians of QFTs.

\section Properties of $\intp$, $\sigma_\pm\argp$, and $\brck{\sta_j}{\sta
_j'}{x}$ 

By \ref(intpd), $\intp F\argp = \intp F(-p)$ and $ (\intp F\argp )^* = \intp
F^*(p)$: if $F(p)$ is a real-valued function of real $p$, then $\intp F(p) \in
\RR$. As the region of integration of the fourfold integral $\intp$ is
symmetric, $\intp F\argp = 0$ if $F\argp$ is odd function of any component of
$p$. Introducing spherical variables so that
$$   p^0 = ir^{1/2} \cos\vartheta \,, \qquad
        \vp = r^{1/2} \sin\vartheta ( \sin\theta \cos\varphi, \sin\theta
	   \sin\varphi, \cos\theta ) \,,                     \eqnm AppA2 $$
we can infer that
$$   \eqalignno{
     \intp f\argpsq &= \pi^2 \int_0^\infty r f(r) \, dr \,,          \cr
     \intp f\argpsq p^\alpha p^\beta &= \OVER 14 \metrictensor^{\alpha\beta}
        \intp \psq\, f\argpsq \,,               &\eqnmm(AppA3)\cr
     \intp f\argpsq p^\alpha p^\beta p^\gamma p^\delta &= \OVER 1{24} 
        [\metrictensor^{\alpha\beta} \metrictensor^{\gamma\delta} 
	+ \metrictensor^{\alpha\gamma} \metrictensor^{\beta\delta} 
	+ \metrictensor^{\alpha\delta} \metrictensor^{\beta\gamma}] 
	\intp \argpsq^2 f\argpsq \,.             \cr}               $$
The reason we defined by \ref(intpd) the invariant, linear functional
$\intp$ on functions $F(p)$ of $p \in \RRRR$ in terms of
$F(ip^0,\vp)$ is that the fourfold integrals over $\RRRR$ of the
functions of $p$ in \ref(AppA3) do not exist.

We use the following properties of $\sigma_\pm\argp$:
$$   \eqalignno{
     \sigma_\pm^\dagger\argp = \sigma_\pm(-p) \,, &\qquad    
       \sigma_\mp^*(p) \sigma^2 = \sigma^2 \sigma_\pm\argp \,,   \cr
     D_\indfer(\Lambda) \sigma_\pm\bigl( \Lambda^{-1} p \bigr) 
     	&= \sigma_\pm\argp D_{-\indfer}(\Lambda) \,,	&\eqnmm(sppro) \cr
     \sigma_+\argp \sigma_-\argp = \sigma_-\argp \sigma_+\argp = -I \,,
     &\qquad \pmatrix{ 0 & \sigma_+\argp \cr -\sigma_-\argp & 0 \cr} 
         = \argpsq^{-1/2} \gamma^\mu p_\mu \,.   \cr}                $$
Hence, under transformations $\sta_{\pm\indfer} \to U(\Lambda, a)
\sta_{\pm\indfer}$ the products $\sigma_\mp\argp \sta_{\pm \indfer}$ transform
as the right-spinor and left-spinor, tt fields $\sta_{\mp \indfer}$ do,
cf.~\ref(Lortr).

For all tt fields $\sta_j$ and $\sta_j'$, $j = 0, \pm 1/2, 1$, and a complex
constant $z$,
$$   \brck{\sta_j}{z\sta_j'}{x} = z\brck{\sta_j}{\sta_j'}{x} \,, \qquad
     \brck{\sta_j}{\sta_j'}{x} = \brck{\sta_j'}{\sta_j}{x}^*,    \eqnm scppr $$
$$   \brck{\sta_j(-p)}{\sta_j'(-p)}{x} = 
	(-1)^{2j} \brck{\sta_j}{\sta_j'}{x} \,,            \eqnm kajje $$
$$   \brck{\sta_j}{i(a\scpr p) \sta_j'}{x} = \brck{i (a\scpr p) 
         \sta_j}{\sta_j'}{x}  \qquad \forall \; a \in \RRRR \,,\eqnm scpr1 $$
$$   \brck{\sta_j}{\sta_j'}{x} = 0 \quad \forall \;\sta_j' \;
         \Longrightarrow \sta_j = 0 \,,         \eqnm scpr3 $$
by \ref(scprd)--\ref(intpd) and \ref(sppro). Hence, $\brck{\sta_j}{\sta_j}{x}$
is a real, but possibly negative, scalar field, and $\brck{\sta_j}{ip
\sta_j}{x}$ is a real, four-vector field. 

\section Asymptotic tt fields

We take scalar fields $\scv{i}\argx$, chiral-bispinor fields $\fsp_m\argx$,
four-vector fields $\ves{i}\argx$ and $A_a\argx$,
symmetric--traceless--second-rank--four-tensor fields $\tes{a}\argx$, and
antisymmetric--second-rank--four-tensor fields $F_a\argx$. And we incorporate
them in the following asymptotic tt fields that satisfy \ref(stlco):
$$   \eqalignno{
     \stlz\argxp &\equiv f_{0i}\argpsq \scs{i}\argx 
        + 2f_{1i}\argpsq \ves{i}\argx \scpr p 
	+ 6 f_{2a}\argpsq p\scpr \tes{a}\argx p \,,         \cr
     \sta_{\pm\indfer\as}\argxp &\equiv [\sigma_-\argp]^{1/2 \pm 1/2}
        \bigl( I , \mp\sigma_+\argp \bigr)  f_{\indfer m}\argpsq 
        \fsp_m\argx \,,                          &\eqnmm(assdf)\cr
     \stlo\argxp &\equiv (1 - \P_l) f_{3a}\argpsq A_a\argx 
	+ \P_l f_{7i}\argpsq \vev{i}\argx        \cr
     &\qquad{}+ [ 2 f_{4a}\argpsq \tev{a}\argx + 2 f_{5a}\argpsq F_a\argx 
        + f_{6i}\argpsq \scv{i}\argx \metrictensor ] p   \cr} $$
(Note, bispinor fields $\fsp_m$ are incorporated in spinor tt fields $\sta_{\pm
\indfer\as}$.). Here: the repeated indices are summed over; $(T a)^\alpha
\equiv T^{\alpha\beta} a_{\beta}$ for a second-rank--four-tensor $T$ and a
four-vector $a$; $(I, \mp\sigma_+ \argp)$ is a $2 \times 4$ matrix, with
$\sigma_\pm \argp$ given by \ref(paufv); $\P_l a \equiv (\psq)^{-1} (p\scpr a)
p$ for a four-vector $a$ [$\,\P_l(1- \P_l)= 0$, and $\P_l F_a\argx p = 0\,$];
$\metrictensor$ is the metric tensor; and $f_{0i}(y)$, $f_{1i}(y)$,
$f_{2a}(y)$, $f_{\indfer m}(y)$, $f_{3a}(y)$, $f_{4a}(y)$, $f_{5a}(y)$,
$f_{6i}(y)$, and $f_{7i}(y)$ are some complex functions of $y \in \RR$;
functions $f_{ui}$ with the same first subscript are assumed to be linearly
independent.

A second-rank four-tensor $T$ can be written as a sum 
$$   T = F + M + \phi \metrictensor \,,                    \eqnm tensy $$
where $F \equiv \half( T - T^t )$ is an antisymmetric tensor; $M \equiv \half (
T + T^t) - \OVER 14 (\Tr T) \metrictensor $ a symmetric-traceless tensor; and
$\phi \equiv \OVER 14 \Tr T$ a scalar ($T^t$ and $\Tr T \equiv T_{\mu\nu}
\metrictensor^{\mu\nu}$ are the transpose and trace of $T$, respectively).
Hence, we may incorporate tensor fields $T_a\argx$ into the four-vector,
asymptotic tt field $\stlo\argxp$, either piecewise, using their parts $F_a
\argx$, $M_a\argx$, and $\phi_a\argx$, or as a whole, choosing $2f_{4a} =
2f_{5a} = f_{6a}$ in \ref(assdf). By \ref(assdf), we can incorporate scalar,
four-vector, and symmetric--second-rank--four-tensor fields into a scalar or a
four-vector asymptotic tt field. However, we cannot incorporate an
antisymmetric-tensor field $F\argx$ into a scalar, asymptotic tt field $\stlz
\argxp$, nor a scalar field and/or a symmetric--second-rank--four-tensor field
into a transversal, four-vector tt field $\stlo\argxp$ such that $\P_l\stlo =
0$. In general, we can incorporate a field that is a direct product of a
totally symmetric, rank-$m$--four-tensor field and of a rank-$n$--four-tensor
(or chiral bispinor) field into an asymptotic, rank-$n$--four-tensor (or
two-component-spinor) tt field.

For the asymptotic tt fields \ref(assdf),
$$   \eqalignno{
     \brck{\stlz}{\stlz}{x} &= \Cc 1 00 ij \scs{i}^* \scs{j} - 
        \Cc 2 11 ij \ves{i}^*\scpr \ves{j}
        + 3 \Cc 3 22 ab \Tr ( \tes{a}^\dagger \tes{b} ) \,,\cr 
     \brck{\sta_{\pm\indfer\as}}{\sta_{\pm\indfer\as}}{x} &= 
        \Cc 1 {\indfer\,\indfer} mn \afsp_m \fsp_n \,,      &\eqnmm(asscp)\cr
     \brck{\stlo}{\stlo}{x} &= \OVER 34 \Cc 1 33 ab A_a^*\scpr A_b 
        - \Cc 2 44 ab \Tr ( \tev{a}^\dagger \tev{b} ) 
	- \Cc 2 55 ab \Tr ( F_{a}^\dagger F_{b} )          \cr
     &\qquad{}- \Cc 2 66 ij \scv{i}^* \scv{j} 
        + \OVER 14 \Cc 1 77 ij \vev{i}^* \scpr \vev{j}
	\,,                                                \cr}        $$
by \ref(scprd)--\ref(intpd) and \ref(AppA3), where $T^\dagger$ is the tensor
adjoint to $T$, and complex constants
$$   \Cc n uv ij \equiv \intp (\psq)^{n-1} f_{ui}^*\argpsq
        f_{vj}\argpsq \,, \qquad n \ge 1 \,.               \eqnm cjdef $$
Let us regard $\Cc n uv ij $ as elements of Hermitian matrices $\CC n uv $,
assume that they are invertible, and denote their inverses by $\CC -n uv $. We
can choose the functions $f_{0i}$, $f_{1i}$, $f_{2a}$, $f_{\indfer m}$,
$f_{3a}$, $f_{4a}$, $f_{5a}$, $f_{6i}$, and $f_{7i}$ in such a way that $\CC 1
00 $, $\CC 2 11 $, $\CC 3 22 $, $\CC 1 {\indfer\,\indfer} $, $\CC 1 33 $, $\CC
2 44 $, $\CC 2 55 $, $\CC 2 66 $, and $\CC 1 77 $ are proportional to identity
matrices (as they are if only one function of each sort is present, since $\CC
n uu $ are then positive numbers); in such a case, there is no interference in
\ref(asscp) between the components of asymptotic tt fields $\stlj\argxp$.

The asymptotic tt fields \ref(assdf) suggest the following fields as potential
macroscopic variables: scalar fields 
$$   \eqalign{
     \scs{i}[x;\sta_0] &\equiv \Cc -1 00 ij \intp f_{0j}^*\argpsq 
        \sta_0\argxp \,,                         \cr
     \scv{i}[x;\sta_1] &\equiv \Cc -2 66 ij \intp f_{6j}^*\argpsq 
        p\scpr \sta_1\argxp \,;                  \cr}      \eqnm mvr01 $$
chiral-bispinor fields
$$   \fsp_m[x;\sta_{\pm\indfer}] \equiv \Cc -1 {\indfer\,\indfer} mn 
        \intp f_{\indfer n}^*\argpsq \pmatrix{\mp I \cr -\sigma_-\argp \cr} 
	[\sigma_+\argp]^{1/2 \pm 1/2}
	\sta_{\pm\indfer}\argxp \,,                        \eqnm mvr02 $$ 
where central, big brackets are $4 \times 2$ matrices; four-vector fields
$$   \eqalign{
     \ves{i}[x;\sta_0] &\equiv 2 \Cc -2 11 ij \intp f_{1j}^*\argpsq
        \sta_0\argxp p \,,                       \cr
     A_a[x;\sta_1] &\equiv \OVER 43 \Cc -1 33 ab \intp f_{3b}^*\argpsq 
        (1 - \P_l) \sta_1\argxp \,,              \cr
     \vev{i}[x;\sta_1] &\equiv 4 \Cc -1 77 ij \intp f_{7j}^*
        \argpsq \P_l \sta_1\argxp \,;            \cr}      \eqnm mvr03 $$
symmetric--traceless--second-rank--four-tensor fields
$$   \eqalignno{
     \tes{a}[x;\sta_0] &\equiv 2 \Cc -3 22 ab \intp f_{2b}^*\argpsq 
       \sta_0\argxp [p \oot p - \OVER 14\psq\, 
       \metrictensor] \,,   				   &\eqnmm(mvr04)\cr
     \tev{a}[x;\sta_1] &\equiv \Cc -2 44 ab \intp f_{4b}^*\argpsq 
       \bigl[ \sta_1\argxp \oot p + p \oot \sta_1\argxp - \half
       p\scpr \sta_1\argxp \metrictensor \bigr] \,;  \cr}  $$
and antisymmetric--second-rank--four-tensor fields
$$   F_a[x;\sta_1] \equiv \Cc -2 55 ab \intp f_{5b}^*\argpsq \bigl[ 
       \sta_1\argxp \oot p - p \oot \sta_1\argxp \bigr] \,.  \eqnm mvr05 $$ 
They satisfy \ref(macas) and \ref(macLt); e.g., $\tes{a}[x; U(\Lambda, a)
\sta_0] = \Lambda^{-1} \tes{a} [ \Lambda x + a; \sta_0] \Lambda $.

For the asymptotic tt fields \ref(assdf), by \ref(scprd)--\ref(intpd) and
\ref(AppA3), the currents corresponding to the global phase invariance
\ref(phsym) are:
$$   \eqalignno{
     &\brck{\stlz}{ip\stlz}{x} = - \Re\{ i \ves{i}^* ( \Cc 2 10 ij \scs{j} 
        + 2 \Cc 3 12 ia \tes{a} ) \} \,,         \cr
     &\brck{\sta_{\pm\indfer\as}}{ip^\mu \sta_{\pm\indfer\as}}{x} = 
         \mp \OVER{i}4 \Cc 3/2 {\indfer\,\indfer} mn \afsp_m
	 \gamma^\mu \fsp_n \,,                   &\eqnmm(asphc)\cr
     &\brck{\stlo}{ip\stlo}{x} = \Re\{ i A_a^* ( \Cc 2 35 ab F_b + \OVER 23
         \Cc 2 34 ab \tev{b} )             
	 + i \vev{i}^* [ \OVER 13 \Cc 2 74 ia \tev{a} + \half \Cc 2 76 ij
         \scv{j} ) \} \,.                        \cr}                $$

The free tt Lagrangians \ref(Lfdef) were defined so that:
$$   \Lfree(\stlz, \sd\stlz) = \Re \{ ( \half \Cc 2 01 ij 
        \scs{i}^* \metrictensor^{\mu\nu} + \Cc 3 21 aj 
        \tes{a}^{*\mu\nu} ) \ldp_\nu \ves{j\mu}  \} \,,    \eqnm asscs $$
$$   \Lfree(\sta_{\pm\indfer\as}, \sd \sta_{\pm\indfer\as}) = 
        \mp \OVER 18 \Cc 3/2 {\indfer\,\indfer} mn	
	   \afsp_m \gamma^\mu \ldp_\mu \fsp_n \,,          \eqnm asscb $$
$$   \displaylines{\qquad
     \Lfree(\stlo, \sd\stlo) = - \Re \{ ( \half \Cc 2 53 ab 
        F_a^{*\mu\nu} + \OVER 13 \Cc 2 43 ab \tev{a}^{*\mu\nu} )
        \ldp_\nu A_{b\mu}              \hfill\cr\hfill
     + ( \OVER 16 \Cc 2 47 aj \tev{a}^{*\mu\nu} 
        + \OVER 14 \Cc 2 67 ij \scv{i}^* \metrictensor^{\mu\nu} ) \ldp_\nu 
	\vev{j\mu}  \} \,,             \qquad\eqnmm(asscv)\cr}         $$
by \ref(Lfdef) and \ref(asphc). If $f_{1i}\argpsq \equiv 0$, then
$\Lfree(\stlz, \sd\stlz) = 0$; if $f_{3a}\argpsq \equiv f_{7i}\argpsq
\equiv 0$, then $\Lfree(\stlo, \sd\stlo) = 0$. As we assumed in
Sec.~2.2 that the interaction part of a tt Lagrangian is strictly
local and contains no derivative couplings, results
\ref(asscs)--\ref(asscv) tell us which combinations of fields and
their derivatives may appear in such a QFT Lagrangian that has a tt
extension \ref(Lttft) with asymptotic tt fields \ref(assdf). Note
that addition of $x$-independent terms to asymptotic tt fields
\ref(assdf) would add only divergences to \ref(asscs)--\ref(asscv).

\section Examples of tt extensions

According to \ref(stjas)--\ref(Lttft), a minimum requirement that a tt
Lagrangian $\Lext$ is a tt extension of a QFT Lagrangian $\Lq$ defined in
terms of certain fields is that (1)~these fields are among the fields that
define the asymptotic tt fields $\stl\argxp$, and (2)~the asymptotic tt
Lagrangian $\Lext(\stl, \sd\stl)$ equals $\Lq$. To get an idea how we can meet
these two conditions, we give a few examples. We will not investigate
alternatives to definition \ref(Llexp) of the part $\L_\lambda(\sta)$ of tt
Lagrangian that determines the asymptotic tt fields $\stl$. 

To consider tt extensions of the Lagrangian for the Dirac equation,
$$   \L_D \equiv -\half \afsp \gamma^\mu \ldp_\mu \fsp
        - m\afsp \fsp \,,                                  \eqnm DQFTL $$
we take a left-spinor tt field $\sta_\indfer\argxp$ and an asymptotic tt field
$\stlh\argxp = (\sigma_-\argp, I) f_\indfer\argpsq \fsp\argx $ with a real
function $f_\indfer\argpsq$. The tt Lagrangians 
$$   \L'_D \equiv 4 \CC -3/2 {\indfer\,\indfer} \Re \brck{
        \sta_\indfer}{\sd\sta_\indfer}{x} - m \CC -1 {\indfer\,\indfer}
	\brck{\sta_\indfer}{\sta_\indfer}{x}               \eqnm DttL1 $$
and
$$   \L''_D \equiv 4 \CC -3/2 {\indfer\,\indfer} \Re \brck
        {\sta_\indfer}{\sd\sta_\indfer}{x} 
	- m \afsp[x; \sta_\indfer] \fsp[x; \sta_\indfer]   \eqnm DttL2 $$
are real, scalar fields and equal $\L_D$ for $\sta = \stl$, by \ref(assdf),
\ref(asscb), \ref(asscp), \ref(macas), and \ref(mvras). If $\CC 3/2
{\indfer\,\indfer} = 4$ and $\CC 1 {\indfer\,\indfer} = m > 0$, then $\L'_D =
\Lfree - \brck{\sta_\indfer}{ \sta_\indfer}{x}$.

There is no direct tt extension of the Lagrangian $\L_{KG} \equiv - | \partial
\phi |^2 - m^2 \phi^2 $ for the Klein-Gordon equation of a real, massive,
scalar field, by \ref(asscs). The Lagrangian
$$   \L_E \equiv -\varphi^\mu \ldp_\mu \phi + \varphi^\mu \varphi_\mu 
	- m^2 \phi^2                                       \eqnm LKGED $$
can be regarded as an equivalent to $\L_{KG}$, because its Euler-Lagrange
equations are equivalent to the Klein-Gordon equation and condition $\varphi =
\partial \phi$. For a real, scalar tt field $\sta_0\argxp$ with $\stlz\argxp =
f_1 \argpsq [ m (\psq)^{1/2} \phi\argx + 2 \varphi\argx\scpr p ]$, we can check
that tt Lagrangians
$$   \L'_E \equiv 2 m^{-1} \CC -5/2 11 \brck{\sta_0}{\sd \sta_0}{x} - 
	\CC -2 11 \brck{\sta_0}{\sta_0}{x}                    \eqnm LKGET $$
and
$$   \L''_E \equiv 2m^{-1} \CC -5/2 11 
        \brck{\sta_0}{\sd \sta_0}{x} + \varphi[x; \sta_0]\scpr 
	\varphi[x; \sta_0] - m^2 \phi^2[x; \sta_0]         \eqnm LKGEt $$
equal $\L_E$ for $\sta_0 = \stlz$. If $\CC 2 11 = \half\, m\CC 5/2 11 = 1$,
then $\L'_E = \Lfree - \brck{\sta_0}{\sta_0}{x}$.

We can generalize the free tt Lagrangian $\brck{\sta_1}{\sd\sta_1}{x}$ of a
real, four-vector tt field $\sta_1\argxp$ as the following sum of two real,
scalar fields:
$$   s_l \brck{\P_l \sta_1}{\sd \P_l\sta_1}{x} +
        s_t \brck{(1-\P_l)\sta_1}{\sd(1-\P_l)\sta_1}{x} \,,   \eqnm gensc $$
where $s_l$ and $s_t$ are two real parameters; for $s_l = s_t = 1$, this
generalization equals $\brck{\sta_1}{\sd \sta_1}{x}$. Generalization
\ref(gensc) with $s_l = 0$ will enable us to construct a local-gauge--invariant
tt Lagrangian that equals for certain asymptotic tt fields the first-order QED
Lagrangian
$$   \L_{QE} \equiv F^{\mu\nu} (\ldp_\nu A_\mu + F_{\mu\nu} )
        - \afsp ( \OVER 12 \gamma^\mu \ldp_\mu 
	+ ie\gamma^\mu A_\mu + m ) \fsp \,.                 \eqnm DQFTL $$
To this end, we choose a complex, left-spinor tt field $\sta_\indfer \argxp$
and a real, four-vector tt field $\sta_1\argxp$; the asymptotic tt fields
$$   \eqalign{
     \stlh\argxp &= f_\indfer\argpsq \bigl( -\sigma_-\argp, I \bigr)
        \fsp\argx \,,                            \cr
     \stlo\argxp &= f_3\argpsq A\argx + 2 f_5\argpsq F\argx p \,;
                                                 \cr}      \eqnm QEDas $$
and fields \ref(mvr02), \ref(mvr05), and
$$   A[x;\sta_1] \equiv \CC -1 33 \intp f_3\argpsq 
        \sta_1\argxp \,.                                   \eqnm TQEDm $$
Take the tt Lagrangian
$$   \eqalign{
     \L'_{QE} &\equiv 2 \CC -2 53 \brck{ (1-\P_l)\sta_1}{\sd 
     	(1-\P_l)\sta_1}{x} + \Tr ( F^\dagger[x; \sta_1] F[x;\sta_1] ) \cr
     &\qquad{}+ 4 \CC -3/2 {\indfer\,\indfer} \Re\brck{\sta_\indfer}{[ 
        \sd - e G\argpsq p\scpr \sta_1 ] \sta_\indfer}{x}    \cr
     &\qquad{}- m \CC -1 {\indfer\,\indfer} 
        \brck{\sta_\indfer}{\sta_\indfer}{x} \,,    \cr}      \eqnm TQFTL $$
where $G\argpsq$ is a real-valued function such that
$$   \eqalign{
     \CC 3/2 {\indfer\,\indfer}  &= \pi^2 \int_0^\infty y^{3/2} 
        f_\indfer^2(-y) f_3(-y) G(-y) \,dy \,,   \cr
     4 \CC 1 33 &= \pi^2 \int_0^\infty y f_3(-y) 
        G^{-1}(-y) \,dy \,.                      \cr}      \eqnm Gcond $$
For the asymptotic tt fields \ref(QEDas) the tt Lagrangian $\L'_{QE}$ equals
$\L_{QE}$. And $\L'_{QE}$ is invariant under the following tt local-gauge 
transformations:
$$   \eqalign{
     \sta_\indfer\argxp &\to e^{ie\alpha\argx} 
        \sta_\indfer\argxp \,,                   \cr
     \sta_1\argxp &\to \sta_1\argxp - G^{-1}\argpsq \P_l 
             \partial \alpha\argx \,.            \cr}      \eqnm TGtra $$
Under tt local-gauge transformations \ref(TGtra), the fields \ref(mvr02),
\ref(mvr05), and \ref(TQEDm) and fields in \ref(QEDas) transform as under the
local-gauge transformations of QED,
$$   \eqalign{
     \fsp[x;\sta_\indfer] &\to e^{ie\alpha\argx}
        \fsp[x;\sta_\indfer] \,,                 \cr
     A[x;\sta_1] &\to A[x;\sta_1] - \partial\alpha\argx \,, \qquad
     F[x;\sta_1] \to F[x;\sta_1] \,.             \cr}      \eqnm QEDGt $$
We do not know how to construct such a gauge-invariant part $\L_\lambda$ of a
tt Lagrangian that in general tends to infinity if $\lambda \to 0$ but for the
asymptotic tt fields \ref(QEDas) that are the most general solution to its
Euler-Lagrange equations, cf.~\ref(Llexp). We could do without a term such as
$\L_\lambda$ in a tt Lagrangian extending $\L_{QE}$, were there a mechanism
inherent to $\L'_{QE}$ that would select the asymptotic tt fields \ref(QEDas)
as the domain of the tt path integral. If so, $\L'_{QE}$ would be a tt
extension of $\L_{QE}$ that is invariant under tt counterpart \ref(TGtra) to
the local-gauge transformations \ref(QEDGt) of QED.

\vfill\eject\bibliography 
Wein4 See, e.g., S.~Weinberg, \it The Quantum Theory of Fields \rm(Cambridge
University Press, Cambridge 1995) Vol.~I, Secs.~12, 1.3, 9.1--9.6, 7.2, 7.3,
and 9; and the references therein. 

Salam A.~Salam, in {\it The Physicist's Conception of Nature}, ed.
J.~Mehra (D.~Reidel, Dordrecht, 1973) p.~430;\\ C.~J.~Isham, A.~Salam and
J.~Strathdee, Phys.\ Rev.\ D\bf 3 \rm (1971) 1805; D\bf 5 \rm (1972) 2548. 

Ross See, e.g., D.~Bailin, Contemp.\ Phys.\ {\bf 30} (1989) 237;\\ D.~J.~Gross,
Nucl.\ Phys.\ B (Proc.\ Suppl.) {\bf15} (1990) 43;\\ G.~G.~Ross, Contemp.\
Phys.\ {\bf 34} (1993) 79;\\ S.~Weinberg, in \it Twentieth Century Physics, \rm
ed. L.~M.~Brown, A.~Pais and B.~Pipard (IOP Publishing and AIP
Publishing, New York, 1995) Vol.~III, p.~2033.

Bjork  J.~D.~Bjorken and S.~D.~Drell, \it Relativistic Quantum 
Fields \rm(McGraw-Hill, New York, 1965) Secs.~11.2 and 11.3;\\ J.~Schwinger,
in \it The Physicist's Conception of Nature, \rm ed. J.~Mehra (D.~Reidel,
Dordrecht, 1973) p.~413.

Heise W.~Heisenberg, Ann.~Phys.~(Leipzig) \bf 32\rm (1938) 20.

Cheng See, e.g., T.~P.~Cheng and L.~F.~Li, \it Gauge Theory of Elementary
Particle Physics, \rm(Claredon Press, Oxford, 1992) Secs.~1.2, 1.1, 5.3, 13.1,
5.1, and~8.1. 

Feynm R.~P.~Feynman, R.~B.~Leighton, and M.~Sands, \it The Feynman Lectures on
Physics \rm(Addison-Wesley, Reading, Mass.,\ 1965) Vol.~II, Sec.~12.7.

Libof See, e.g., R.~L.~Liboff, \it Kinetic Theory \rm(Prentice-Hall, Englewood
Cliffs, N.~J., 1990) Ch.~3 and Sec.~2.4.

Willi See, e.g., M.\ M.\ R.\ Williams, \it Mathematical Methods in Particle 
Transport Theory, \rm(Butterworths, London, 1971) Sec.~2.7 and Ch.~11.

Groot See, e.g., S.~R.~de Groot, W.~A.~van Leeuwen and Ch.~G.~van Weert, 
\it Relativistic Kinetic Theory \rm(North-Holland, Amsterdam, 1980)
Secs.~I.2, VI.1 and VII.

Grad See, e.g., H.~Grad, in \it Application of Nonlinear Partial
Differential Equations in Mathematical Physics, \rm Proc.\ Symp.\ Appl.\ Math.\
{\bf XVII} (Am.\ Math.\ Soc., Providence, R.~I., 1965) p.~154, and
references therein.

Wein1 See, e.g., S.~Weinberg, \it Dreams of a Final Theory \rm(Pantheon Books,
New York 1992).

mi004 M.~Ribari\v c and L.~\v Su\v ster\v si\v c, Transp.\ Theory Stat.\ Phys.\
\bf 24\rm (1995) 1. 

mi002 M.~Ribari\v c and L.~\v Su\v ster\v si\v c, Int.\ J.\ Theor.\ Phys.\ 
\bf 34\rm (1995) 571.

Weidn R.~T.~Weidner, in \it The New Encyclopaedia Britannica \rm (Encyclopaedia
Britannica, Chicago, 1986) 15th edition, Vol.~25, p.845.

Wein2 See, e.g., S.~Weinberg, Phys.\ Rev.\ D\bf7\rm (1973) 1068, Sec.~II;
Rev.\ Mod.\ Phys.\ {\bf 46} (1974) 255, Secs.~I and~III.

Schw1 J.~Schwinger, Phys.\ Rev.\ \bf 91\rm (1953) 713;\\ R.~L.~Arnowitt and
S.~I.~Fickler, Phys.\ Rev.\ \bf 127\rm (1962) 1821.

Fadee See, e.g., L.~D.~Faddeev and A.~A.~Slavnov, \it Gauge Fields:
Introduction to Quantum Theory \rm(Benjamin Cummings, Reading, Mass., 1980)
2nd edition, Sec.~3.2. 

mi006 For an example see M.~Ribari\v c and L.~\v Su\v ster\v si\v c, Transp.\
Theory Stat.\ Phys.\ \bf 16 (1987) 1041, Secs.~4.5 and 5.1.

Wein3 S.~Weinberg, Ann.\ Phys.\ (N.Y.) {\bf 194} (1989) 336. 

Sudbe For an introduction see, e.g., A.~Sudbery, \it Quantum Mechanics and
Particles of Nature \rm(Cambridge University Press, Cambridge 1988) Chs.~7, 5,
and 2.

Frogg See, e.g., C.~D.~Froggatt and H.~B.~Nielsen, \it Origin of Symmetries,
\rm(World Scientific, Singapore, 1991), and references therein.

McKeo See, e.g., J.~Bernstein, Rev.\ Mod. Phys.\ \bf 46\rm (1974) 7, footnote
32;\\ D.~G.~C.\ McKeon, Can.\ J.\ Phys.\ {\bf 72} (1994) 601.

Grein W.~Greiner, \it Quantum Mechanics, an Introduction, \rm(Springer Verlag,
Berlin, 1989) Sec.~13.2.

Feyng For some related comments see K.~Gottfried and V.~F.~Weisskopf, \it
Concepts of Particle Physics \rm(Claredon Press, Oxford 1984) Vol.~I,
Sec.~13c;\\ R.~P.~Feynman, F.~B.~Morinigo, and W.~G.~Wagner, \it Feynman
Lectures on Gravitation, \rm ed. B.~Hatfield (Addison-Wesley, Reading,
Mass., 1995) Secs~1.4 and 1.5;\\ I.~Percival, Phys.~World \bf 10 
(1997) (3) 43, and references therein.

mi001 See, e.g., M.~Ribari\v c and L.~\v Su\v ster\v si\v c, Fizika B \bf 3,
\rm (1994) 93; Found.\ Phys.\ Lett.\ {\bf 7} (1994) 531; and references
therein.

Chiao See, e.g., R.~Y.~Chiao, P.~G.~Kwiat, and A.~E.~Steinberg, Sci.\ Am.\ \bf
269 (1993) \rm(2), 38;\\ R.~Y.~Chiao, J.~Boyce, and J.~C.~Garrison, in \it
Fundamental Problems in Quantum Theory, \rm ed. D.~G.~Greenberger and
A.~Zeilinger (The New York Academy of Sciences, New York, 1995) p.~400, and
references therein.

Polya A.~M.~Polyakov, \it Gauge Fields and Strings \rm(Harwood Academic
Publishers, Chur, 1987) Sec.~1.2.

Kinos T.~Kinoshita and W.~J.~Marciano, in \it Quantum Electrodynamics, \rm
ed. T.~Kinoshita (World Scientific, Singapore, 1990) p.~419.

Brods S.~J.~Brodsky, Preprint SLAC-PUB-95-6781, hep-ph/9503391.

\relax

\bye